\documentclass[11pt,prd,a4paper,preprintnumbers,amsmath,amssymb,nofootinbib]{article}
\pdfoutput=1
\usepackage{feynmp-auto,expdlist}
\usepackage{amsmath,amsfonts,amssymb}
\usepackage{graphicx}
\usepackage{caption}
\usepackage{enumerate}
\usepackage{hyperref}
\usepackage{latexsym}
\usepackage{hepnicenames}
\usepackage{enumerate}
\usepackage{soul}
\usepackage[normalem]{ulem}
\usepackage{wasysym}
\usepackage{makecell}
\usepackage{bbm}

\font\tenrsfs=rsfs10 at 12pt
\font\sevenrsfs=rsfs7
\font\fiversfs=rsfs5
\newfam\rsfsfam
\textfont\rsfsfam=\tenrsfs
\scriptfont\rsfsfam=\sevenrsfs
\scriptscriptfont\rsfsfam=\fiversfs

\numberwithin{equation}{section}

\usepackage{mathrsfs}
\usepackage{braket}
\usepackage{titling}
\usepackage{amsmath}
\usepackage{slashed}
\usepackage{amssymb}
\usepackage{epsfig}
\usepackage{graphicx}
\usepackage{color}
\usepackage{rotating}
\usepackage{hyperref}
\usepackage[margin=1.in]{geometry}
\usepackage[table,xcdraw,dvipsnames]{xcolor}
\usepackage[compress,numbers,sort]{natbib}
\usepackage{colortbl}
\usepackage{pdflscape}
\usepackage{color}
\usepackage{mathtools}
\usepackage{colortbl}
\usepackage{comment}

\newcommand{\X}{{\cal X}}

\newcommand{\SU}{{\rm SU}}

\newcommand{\U}{{\rm U}}

\definecolor{nicered}{rgb}{0.7,0.1,0.1}
\definecolor{nicegreen}{rgb}{0.1,0.5,0.1}
\definecolor{red}{rgb}{1.0, 0, 0}
\definecolor{niceblue}{rgb}{0,0,0.8}
\allowdisplaybreaks

\definecolor{blus}{cmyk}{1,1,0,0.6}
\definecolor{verde}{cmyk}{0.92,0,0.59,0.25}
\definecolor{rossos}{cmyk}{0,1,1,0.55}
\definecolor{nicepurple}{RGB}{193,67,213}

\hypersetup{colorlinks,bookmarksopen,bookmarksnumbered,linkcolor=blus,pdfstartview=FitH,urlcolor=rossos,citecolor=verde}

\def\eq#1{{Eq.~(\ref{#1})}}
\def\eqs#1#2{{Eqs.~(\ref{#1})--(\ref{#2})}}
\def\fig#1{{Fig.~\ref{#1}}}
\def\figs#1#2{{Figs.~\ref{#1}--\ref{#2}}}
\def\Table#1{{Table~\ref{#1}}}

\def\sect#1{{Section~\ref{#1}}}

\def\app#1{{Appendix~\ref{#1}}}
\def\apps#1#2{{Appendices~\ref{#1}--\ref{#2}}}
\def\Im{\mbox{Im}\,}
\def\Re{\mbox{Re}\,}

\def\diag{\mbox{diag}\,}
\renewcommand{\bar}{\overline}

\newcommand{\beq}{\begin{equation}}
\newcommand{\eeq}{\end{equation}}
\newcommand{\bea}{\begin{eqnarray}}
\newcommand{\eea}{\end{eqnarray}}

\renewcommand{\[}{\left[}
\renewcommand{\]}{\right]}
\renewcommand{\(}{\left(}
\renewcommand{\)}{\right)}

\renewcommand{\X}{\mathcal{X}}

\def\be{\begin{equation}}
\def\ee{\end{equation}}

\newcommand{\Kmix}{$K^0 - \bar K^0$ }
\newcommand{\Bdmix}{$B_d^0 - \bar B_d^0$ }
\newcommand{\Bsmix}{$B_s^0 - \bar B_s^0$ }
\newcommand{\Dmix}{$D^0 - \bar D^0$ }

\begin{document}

\begin{center}  
{\Huge
\bf\color{blus} 
On the 
IR/UV flavour connection 
\\ \vspace{0.3cm} 
in non-universal 
axion models 
} \\
\vspace{0.8cm}

{\bf Luca Di Luzio, Alfredo Walter Mario Guerrera, \\ Xavier Ponce D\'iaz, Stefano Rigolin}\\[7mm]

{\it Istituto Nazionale di Fisica Nucleare (INFN), Sezione di Padova, \\
Via F. Marzolo 8, 35131 Padova, Italy}\\[1mm]
{\it Dipartimento di Fisica e Astronomia `G.~Galilei', Universit\`a di Padova,
 \\ Via F. Marzolo 8, 35131 Padova, Italy
}\\[1mm]

\vspace{0.3cm}

\begin{quote}

Non-universal axion models, with the Peccei-Quinn (PQ) symmetry acting on Standard Model (SM) fermions in a 
generation-dependent way, are typically accompanied by two different sources of flavour violation, dubbed 
here as infrared (IR) and ultraviolet (UV). The former is due to the flavour violating axion couplings 
to SM fermions, while the latter arises from the heavy degrees of freedom that UV complete the axion 
effective field theory. We point out that these two sources of flavour violation are directly related 
and exemplify this connection in a general class of non-universal axion model, based on a renormalizable 
DFSZ-like setup with two Higgs doublets (PQ-2HDM). We next discuss the interplay of axion flavour 
phenomenology with the signatures stemming from the heavy radial modes of the PQ-2HDM, including 
meson oscillation observables and charged lepton flavour violating decays.
We emphasize the strong complementarity between flavour observables, LHC direct searches and standard axion physics.

\end{quote}

\thispagestyle{empty}

\end{center}

\bigskip
\tableofcontents

\section{Introduction}

The Peccei-Quinn (PQ) solution to the strong CP problem \cite{Peccei:1977hh,Peccei:1977ur} has emerged as a well-motivated 
paradigm beyond the Standard Model (SM), by delivering in the low-energy effective field theory (EFT) the axion field 
\cite{Weinberg:1977ma,Wilczek:1977pj} which also provides a natural dark matter candidate \cite{Dine:1982ah,Abbott:1982af,
Preskill:1982cy}. Invisible axion models can be classified in two broad categories, known conventionally as DFSZ 
\cite{Zhitnitsky:1980tq,Dine:1981rt} and KSVZ \cite{Kim:1979if,Shifman:1979if} models, depending on whether the SM 
fermions are charged or not under the $\U(1)_{\rm PQ}$ symmetry. In the former case, only the scalar sector of the SM 
is extended, by a second Higgs doublet and a SM singlet that allows to establish the hierarchy between the PQ and 
electroweak breaking scales. In the original formulation of the DFSZ model \cite{Zhitnitsky:1980tq,Dine:1981rt} the 
SM fermions PQ-charge assignment is flavour blind, namely SM fermions with the same hypercharge share also the same PQ 
charge. On the other hand, non-universal variants of the DFSZ model (and in particular of the original WW axion model  \cite{Weinberg:1977ma,Wilczek:1977pj}), in which the PQ symmetry acts on SM fermions in a generation-dependent way, 
appeared in the late 80's \cite{Peccei:1986pn,Krauss:1986wx,Krauss:1987ud}, in the effort to explain in terms of a 
heavy axion the anomalous observation of a narrow $e^+e^-$ pair peak at 1.8 MeV in heavy ion collision at the GSI 
\cite{Clemente:1983qh,Schweppe:1983yv}. 

From a modern perspective, non-universal axion models can be motivated in several respects: $i)$ their potential connection to 
the SM flavour puzzle, namely explaining the observed pattern of SM fermion masses and mixings in terms of a non-universal 
$\U(1)_{\rm PQ}$ acting as a flavour 
symmetry
(see e.g.~\cite{Ema:2016ops,Calibbi:2016hwq,Arias-Aragon:2017eww,Bjorkeroth:2018ipq}); 
$ii)$ the possibility of suppressing the axion couplings to nucleons and electrons, leading to the so-called ``astrophobic'' axion 
scenario \cite{DiLuzio:2017ogq} (see also \cite{Bjorkeroth:2019jtx,Darme:2020gyx,DiLuzio:2022tyc,Badziak:2023fsc,Takahashi:2023vhv}),  
which allows to relax astrophysical bounds on the axion decay constant;
$iii)$ the experimental opportunity of discovering the axion via flavoured axion searches (see e.g.~\cite{Cornella:2019uxs, 
MartinCamalich:2020dfe,Calibbi:2020jvd,Bauer:2021mvw,Guerrera:2021yss,Gallo:2021ame,Jho:2022snj,Guerrera:2022ykl}), which offer a complementary 
probe with respect to standard axion detection strategies \cite{Irastorza:2018dyq,Sikivie:2020zpn}.

One of the most striking consequences of non-universal axion models is that, after going to the mass basis, fermion rotation 
matrices beyond the CKM/PMNS combinations become physical and thus can be probed via flavour violating axion couplings to SM 
fermions. We will indicate this type of 
flavour violation as of low-energy or infrared (IR) type, in the sense that it stems from the light axion field. However, this 
is not the only flavour violation to be expected, since the ultraviolet (UV) completion of the axion EFT might leave behind 
extra sources of flavour violation, indicated in the following as of high-energy or 
UV type. While we expect 
this to be a generic feature of axion EFTs, the scope of the present paper is to exemplify this connection in a general class 
of renormalizable (i.e.~fully calculable) non-universal DFSZ-like models, whose scalar sector is based on a SM singlet and two 
Higgs doublets which couple non-universally to different generations of up- and down-type quarks (and charged leptons). 
Since the SM fermion mass matrices and the Yukawa interactions of the two Higgs doublet model (2HDM) are non-simultaneously 
diagonalizable, new flavour violating UV sources arise from the Yukawa couplings of the 2HDM degrees of freedom. Note that 
the mass scale of the non-SM-like Higgs scalars is a free parameter, ranging from a few hundreds of GeV (experimentally 
bounded by LHC direct searches) up to the PQ breaking scale, of the order of the axion decay constant, 
$f_a$. In this paper, we will entertain the bottom-up perspective that the 2HDM is not completely decoupled from the 
TeV scale,\footnote{A low-scale 2HDM within a PQ setup is theoretically motivated by the possibility of establishing 
the hierarchy between the electroweak and PQ scales in a ``technically natural way'', i.e.~stable 
under radiative corrections thanks to the presence of ``ultra-weak'' couplings in the scalar potential (see 
\app{app:2HDMpotential} for further details).} and thus its degrees of freedom can be directly tested at the LHC and 
indirectly via flavour violating processes. 

The aim of this paper, is to underline the general correlation between IR and UV sources of flavour violation in non-universal 
DFSZ-like axion models, focusing on flavour violating observables both in the quark and lepton sector.\footnote{For a related 
work focusing on flavour violation in the charged-lepton sector, see Ref.~\cite{Badziak:2021apn}.} In particular, we 
will show that the pattern of flavour violation involving the light axion field is directly correlated to that of the 
heavy radial modes of the 2HDM. Assuming for instance a QCD axion discovery in a golden channel like e.g.~$K \to \pi a$ 
(which would definitely point to a tree-level source of flavour violation in $s$-$d$ axion couplings), allows us to 
predict a definite pattern of flavour violation from the 2HDM heavy scalars in $s$-$d$ transitions, e.g.~in $K^0$-$\bar K^0$ 
mixing. The bound arising from the latter observable can be then confronted with LHC direct searches, as well as with 
astrophysical bounds on the axion decay constant. Similar considerations apply also to other phenomenologically relevant 
quark flavour transitions involving $B$ and $D$ mesons or to lepton flavour violating (LFV) decays. 

The paper is structured as follows. \sect{sec:nonuniversalaxion} focuses on theoretical aspects of the IR/UV flavour 
connection. We first introduce a general class of non-universal axion models (the only assumption being here a minimal 
field content and renormalizability) and recall the pattern of flavour violating axion couplings in terms of PQ charges. 
Next, we discuss the structure of a general flavour non-conserving 2HDM in the presence of a $\U(1)_{\rm PQ}$ symmetry 
(hereby denoted as PQ-2HDM),  and work out the flavour violating couplings of the 2HDM scalar fields and  the associated 
4-fermion operators that are generated after integrating out the heavy radial modes. A simplified class of non-universal 
axion models is also introduced, in order to have an explicit realization as a benchmark scenario. Phenomenological 
aspects, including LHC bounds and flavour constraints on the non-universal 2HDM are discussed in \sect{sec:pheno}, 
together with the aforementioned interplay between flavoured axion searches and other more standard flavour violating 
processes, both in the hadronic and leptonic sectors. We summarize our conclusions in \sect{sec:concl}, while more 
technical details about the scalar potential and flavour violating observables are deferred to \apps{sec:scalpot}{app:LFVobs}.

%%%%%%%%%%%%%%%%%%%%%%%%%%%%%%%%%%%%%%%%%%%%%%%%%%%%%%%%%%%%%%%%%%%%%%%%%%%%%%%%%%%%%%%%%%%%%%%%%%%%%%%%%%%%%%%%%%%
%%%%%%%%%%%%%%%%%%%%%%%%%%%%%%%%%%%%%%%%%%%%%%%%%%%%%%%%%%%%%%%%%%%%%%%%%%%%%%%%%%%%%%%%%%%%%%%%%%%%%%%%%%%%%%%%%%%

\section{Non-universal DFSZ axion models} 
\label{sec:nonuniversalaxion}

%%%%%%%%%%%%%%%%%%%%%%%%%%%%%%%%%%%%%%%%%%%%%%%%%%%%%%%%%%%%%%%%%%%%%%%%%%%%%%%%%%%%%%%%%%%%%%%%%%%%%%%%%%%%%%%%%%%
%%%%%%%%%%%%%%%%%%%%%%%%%%%%%%%%%%%%%%%%%%%%%%%%%%%%%%%%%%%%%%%%%%%%%%%%%%%%%%%%%%%%%%%%%%%%%%%%%%%%%%%%%%%%%%%%%%%

The scalar sector of the DFSZ model includes two Higgs doublets $H_{1,2} \sim (1,2,-1/2)$ endowed with PQ charges 
$\X_{1,2}$ and an extra complex SM-singlet $\phi\sim (1,1,0)$, with PQ charge $\X_\phi$, 
needed for decoupling the PQ and electroweak breaking scales. Singlet and 
doublet scalar fields are coupled via a single non-hermitian operator, $H_2^\dag H_1 \phi$, thus implying $\X_2 - \X_1 = 
\X_\phi = 1$, by normalizing to 1 the PQ charge of the singlet.\footnote{Other choices, e.g.~$H_2^\dag H_1\phi^2$ as 
in the original DFSZ model \cite{Zhitnitsky:1980tq,Dine:1981rt}, are possible as well at the renormalizable level. 
The general conclusions of this paper do not depend on this particular choice, which affects instead the periodicity 
of the axion potential via the so-called domain-wall number (see e.g.~\cite{DiLuzio:2020wdo}).} 
In the following, we will also require the orthogonality of the PQ and hypercharge currents (see e.g.~\cite{DiLuzio:2020wdo}), 
thus ensuring the absence of a mixed  $Z^\mu \partial_\mu a$ kinetic term. This fixes the PQ charges of the two Higgs doublets 
as $\X_1 = -s^2_\beta \equiv - \sin^2\beta$ and $\X_2 = c^2_\beta \equiv \cos^2\beta$, in terms of the vacuum angle, 
$t_\beta \equiv \tan\beta = v_2 / v_1$, where $v_{1,2}$ are the $H_{1,2}$ vacuum expectation values (VEVs) with $v^2 = 
v_1^2 + v_2^2 \simeq (246 \ \text{GeV})^2$. Other details concerning the renormalizable DFSZ scalar potential, 
the implementation of the hierarchy between the PQ and the electroweak scales and the resulting scalar spectrum are 
deferred to \app{sec:scalpot}. 

The Yukawa sector of the 2HDM in the presence of a $\U(1)_{\rm PQ}$ symmetry can be written in a compact form as
\begin{align} 
\label{eq:Lyuk2HDM}
\mathcal{L}_Y^{\rm PQ\text{-}2HDM} = 
&- \bar q_L (Y_1^u H_1 + Y_2^u H_2 ) u_R
+ \bar q_L (Y_1^d \tilde H_1 + Y_2^d \tilde H_2 ) d_R \nonumber \\
&+ \bar \ell_L ( Y_1^e \tilde H_1 + Y_2^e \tilde H_2 ) e_R 
+ \text{h.c.} \, , 
\end{align}
where $\tilde H_{1,2} = (i \sigma^2) H_{1,2}^*$ and $Y^{f}_{1,2}$ are generic three-dimensional matrices. We do not 
specify the origin of neutrino masses, the effects of which will be neglected in the present discussion.
Requiring the $\U(1)_{\rm PQ}$ invariance on the Yukawa Lagrangian implies the following sum rules in flavour space: 
\begin{align}
\label{eq:chargesu}
-\X_q Y_{1,2}^u + Y_{1,2}^u \X_u + \X_{1,2} Y_{1,2}^u &= 0 \, , \\
\label{eq:chargesd}
-\X_q Y_{1,2}^d + Y_{1,2}^d \X_d - \X_{1,2} Y_{1,2}^d &= 0 \, , \\
\label{eq:chargese}
-\X_\ell Y_{1,2}^e + Y_{1,2}^e \X_e - \X_{1,2} Y_{1,2}^e &= 0 \, , 
\end{align}
where the left ($\X_{q,\ell}$) and right ($\X_{u,d,e}$) SM fermion PQ charges are diagonal, but in general not universal, 
three-dimensional matrices. Note that the PQ symmetry requires some of the Yukawa entries to vanish. For instance, in the 
universal DFSZ framework \cite{Zhitnitsky:1980tq,Dine:1981rt} one imposes $Y_1^u = Y_2^{d,e} = 0$, falling into a 
type-II 2HDM scenario. In \sect{sec:textures} we will provide two explicit examples of non-universal PQ-2HDM Yukawa textures. 

%%%%%%%%%%%%%%%%%%%%%%%%%%%%%%%%%%%%%%%%%%%%%%%%%%%%%%%%%%%%%%%%%%%%%%%%%%%%%%%%%%%%%%%%%%%%%%%%%%%%%%%%%%%%%%%%%%%
\subsection{Flavour violation from the light axion field (IR source)}
%%%%%%%%%%%%%%%%%%%%%%%%%%%%%%%%%%%%%%%%%%%%%%%%%%%%%%%%%%%%%%%%%%%%%%%%%%%%%%%%%%%%%%%%%%%%%%%%%%%%%%%%%%%%%%%%%%%

The low-energy axion field, which spans over the neutral angular modes of the scalar multiplets
\beq 
\phi \supset \frac{v_\phi}{\sqrt{2}} e^{i a_\phi / v_\phi} \, ,  \quad 
H_{1} \supset \frac{v_{1}}{\sqrt{2}} e^{i a_{1} / v_{1}} 
\begin{pmatrix}
1 \\ 
0 
\end{pmatrix} \, , 
\quad 
H_{2} \supset \frac{v_{2}}{\sqrt{2}} e^{i a_{2} / v_{2}} 
\begin{pmatrix}
1 \\ 
0 
\end{pmatrix} \, ,
\eeq
is defined as \cite{DiLuzio:2020wdo} 
\beq 
a = \frac{1}{v_{\rm PQ}} \left(v_\phi a_\phi + v (-s^2_\beta c_\beta a_1 + c^2_\beta s_\beta a_2) \right) 
\simeq a_\phi \, , \qquad v_{\rm PQ}^2 = v_\phi^2 + v^2 c^2_\beta s^2_\beta \simeq v_\phi^2 \, , 
\eeq
where the $\simeq$ relation holds once $v_{\rm \phi} \gg v$ is imposed. The axion effective Lagrangian below the electroweak scale reads \cite{DiLuzio:2020wdo} 
\beq
\label{eq:LeffEW}
\mathcal{L}_a \supset 
\frac{g_s^2}{32\pi^2} \frac{a}{f_a} G \tilde G \,+\, \frac{E}{N} \frac{e^2}{32\pi^2} \frac{a}{f_a} F \tilde F \,+ \!\!
\sum_{f=u,d,e}\frac{\partial_\mu a}{2 f_a} \bar f_i\gamma^\mu\left((C^V_f)_{ij}+(C^A_f)_{ij}\gamma_5 \right) f_j \, , 
\eeq
where $E/N$ 
is the ratio between the electromagnetic and QCD anomaly 
of the PQ current (for typical values in 
concrete axion models, see e.g.~\cite{DiLuzio:2016sbl,DiLuzio:2017pfr}), 
and the axion decay constant is defined by $f_a = v_{\rm PQ} / (2N)$. 
The axion vector and axial couplings to fermions can be written as
\beq 
\label{eq:CRLf}
C^{V,A}_f =  C^R_f \pm C^L_f \, , \qquad
C^{R,L}_f = \frac{1}{2N} V_{f_{R,L}}^\dag \X_{f_{R,L}} V_{f_{R,L}} \, \qquad (f = u,d,e) \,,
\eeq
in terms of the PQ charges $\X_{f_{L,R}}$ and of the unitary matrices, $V_{f_{L,R}}$, which diagonalize the SM fermion mass terms: 
\beq 
\hat{M}_f \equiv V^\dag_{f_L} M_f V_{f_R} \,.
\eeq   
Therefore, the $C_f^{V,A}$ parameters encode all the IR sources of flavour violation, stemming from the light axion field. 
In the present work we will neglect renormalization group corrections on the $C_f^{V,A}$ couplings (see \cite{Choi:2017gpf,Chala:2020wvs,Bauer:2020jbp,Choi:2021kuy,DiLuzio:2022tyc}). This is a reasonable approximation under 
the assumption of a (at most) TeV-scale 2HDM, being running effects proportional to $\log m_H / m_t$, with $m_H$ 
labelling the typical mass scale of the 2HDM radial modes in the decoupling limit (see \app{sec:scalpot} for a 
detailed discussion of the scalar spectrum).

%%%%%%%%%%%%%%%%%%%%%%%%%%%%%%%%%%%%%%%%%%%%%%%%%%%%%%%%%%%%%%%%%%%%%%%%%%%%%%%%%%%%%%%%%%%%%%%%%%%%%%%%%%%%%%%%%%%
\subsection{Flavour violation from the heavy radial modes (UV source)} 
\label{sec:PQ2HDM}
%%%%%%%%%%%%%%%%%%%%%%%%%%%%%%%%%%%%%%%%%%%%%%%%%%%%%%%%%%%%%%%%%%%%%%%%%%%%%%%%%%%%%%%%%%%%%%%%%%%%%%%%%%%%%%%%%%%

The SM fermion mass matrices in the PQ-2HDM follow from \eq{eq:Lyuk2HDM} after setting the Higgs fields on their VEVs, and read 
\begin{align}
\label{eq:defMYuk}
M_u = \frac{v}{\sqrt{2}} c_\beta Y^u_1 + \frac{v}{\sqrt{2}} s_\beta Y^u_2 \, , \ \
M_d = \frac{v}{\sqrt{2}} c_\beta Y^d_1 + \frac{v}{\sqrt{2}} s_\beta Y^d_2 \, , \ \
M_e = \frac{v}{\sqrt{2}} c_\beta Y^e_1 + \frac{v}{\sqrt{2}} s_\beta Y^e_2 \, .  
\end{align}
Since in non-universal PQ models the mass matrices and the Yukawa interactions are not simultaneously diagonalizable, 
flavour violating Higgs couplings to fermions are generated at tree level both in the neutral and charged sector. To derive 
those interactions, we first decompose the two Higgs doublets in terms of mass and charge eigenstates, comprising $h$ (the SM-like 
Higgs) and $H,A,H^+$ (the heavy radial modes). In the unitary gauge one has  (see \app{app:2HDMpotential}) 
\beq 
\label{eq:PQHiggsdoublets}
H_1 = 
\begin{pmatrix}
\frac{1}{\sqrt{2}} (c_\beta v - s_\alpha h + c_\alpha H  - i s_\beta A ) \\
- s_\beta H^- 
\end{pmatrix} \, , \quad  
H_2 = 
\begin{pmatrix}
\frac{1}{\sqrt{2}} (s_\beta v + c_\alpha h + s_\alpha H  + i c_\beta A ) \\
c_\beta H^-
\end{pmatrix} \, .
\eeq
After unfolding the $\SU(2)_L$ structure in \eq{eq:Lyuk2HDM} and projecting onto the fermion mass basis, 
$f_{L,R} \to V_{f_{L,R}} f_{L,R}$ (with $f = u,d,e$), we obtain the following Yukawa Lagrangian 
\begin{align} 
 \label{eq:LyukM1v2}
\mathcal{L}_Y^{\rm PQ\text{-}2HDM} \supset   
&-\sum_{f=u,d,e} \bar f_{Li} C^{h_f}_{ij} f_{Rj} h 
-\sum_{f=u,d,e} \bar f_{Li} C^{H_f}_{ij} f_{Rj} H
-\sum_{f=u,d,e} \bar f_{Li} C^{A_f}_{ij} f_{Rj} A \\
& - \sqrt{2} \left( \bar d_{Li}   V^\dag_{ik} C^{H^-_u}_{kj} u_{Rj} H^-  
    + \bar u_{Li} V_{ik} C^{H^+_d}_{kj} d_{Rj} H^+ 
    + \bar \nu_{Li} U^\dag_{ik} C^{H^+_e}_{kj} e_{Rj} H^+  \right)+ \text{h.c.} \, , \nonumber
\end{align} 
where the CKM and PMNS mixing matrix entering in the charged-current interactions are defined respectively as $V \equiv V_{\rm CKM} 
= V^\dag_{u_L} V_{d_L}$ and $U \equiv U_{\rm PMNS} = V^\dag_{e_L} V_{\nu_L}$, and the Higgs 
flavour violating coefficients $C_{ij}$ read 
(see also \cite{Badziak:2021apn})
\begin{align}
C^{h_u}_{ij} &= \frac{m_{u_i}}{v s_\beta} \delta_{ij} c_\alpha + \frac{1}{\sqrt{2}} \frac{-c_{\alpha-\beta}}{s_\beta} \epsilon^u_{ij} \, , 
\qquad\,\,
C^{h_d(h_e)}_{ij} = \frac{m_{d_i(e_i)}}{v c_\beta} \delta_{ij} (- s_\alpha) + \frac{1}{\sqrt{2}} \frac{c_{\alpha-\beta}}{c_\beta} \epsilon^{d(e)}_{ij} \, ,  \label{eq:wilson_coef2} \\
C^{H_u}_{ij} &= \frac{m_{u_i}}{v s_\beta} \delta_{ij} s_\alpha + \frac{1}{\sqrt{2}} \frac{-s_{\alpha-\beta}}{s_\beta} \epsilon^u_{ij} \, , \quad\ \ \,
%\label{eq:wilson_coef4} \, , \\
C^{H_d(H_e)}_{ij} = \frac{m_{d_i(e_i)}}{v c_\beta} \delta_{ij} c_\alpha + \frac{1}{\sqrt{2}} \frac{s_{\alpha-\beta}}{c_\beta} \epsilon^{d(e)}_{ij}  \label{eq:wilson_coef5} \, , \\
C^{A_u}_{ij} &= \frac{m_{u_i}}{v s_\beta} \delta_{ij} (i c_\beta) + \frac{1}{\sqrt{2}} \frac{-i}{s_\beta} \epsilon^u_{ij} \, ,  \qquad\ \
C^{A_d(A_e)}_{ij} = \frac{m_{d_i(e_i)}}{v c_\beta} \delta_{ij} (i s_\beta) + \frac{1}{\sqrt{2}} \frac{-i}{c_\beta} \epsilon^{d(e)}_{ij} \label{eq:wilson_coef8}  \, , \\
C^{H^-_u}_{ij} & = \frac{m_{u_i}}{v s_\beta} \delta_{ij} c_\beta - \frac{1}{\sqrt{2}} \frac{\epsilon^u_{ij}}{s_\beta}  \, ,  \qquad\quad\ \ \ 
C^{H^+_d(H^+_e)}_{ij} = \frac{m_{d_i(e_i)}}{v c_\beta} \delta_{ij} s_\beta - \frac{1}{\sqrt{2}} \frac{\epsilon^{d(e)}_{ij}}{c_\beta}  \label{eq:wilson_coef11} \, , 
\end{align} 
with 
\beq
\label{eq:EpsilonCouplings}
\epsilon^u_{ij} = (V^\dag_{u_L} Y_1^u V_{u_R})_{ij} \, , \qquad 
\epsilon^d_{ij} = (V^\dag_{d_L} Y_2^d V_{d_R})_{ij} \, , \qquad 
\epsilon^e_{ij} = (V^\dag_{e_L} Y_2^e V_{e_R})_{ij} \, . 
\eeq 
Therefore, all the neutral and charged flavour violating sources in the Higgs sector are encoded into the $\epsilon^{f}$ 
parameters. Note that, besides the heavy Higgs radial modes, also the SM-like Higgs features flavour violating couplings 
(cf.~\eq{eq:wilson_coef2}). However, all the tree-level flavour changing neutral current contributions arising from the 
light Higgs sector decouple faster than the radial mode ones, since $c_{\alpha - \beta} \propto 1 / m_H^2$ in the decoupling 
limit (cf.~\eq{eq:cos2alphabeta}).

To connect with typical low-energy flavour observables, like for example meson oscillations, one has to integrate out both the heavy radial 
modes, $H,A,H^+$ and the SM--like Higgs, $h$, obtaining the $d=6$ low-energy effective Lagrangian describing the (light) quarks 
and leptons 4-fermion interactions
\beq 
\mathcal{L}^{\rm 4\text{-}fermion}_{\rm EFT} = \mathcal{L}^{\rm 4\text{-}quark}_{\rm EFT} 
+ \mathcal{L}^{\rm semi\text{-}lept}_{\rm EFT} 
+  \mathcal{L}^{\rm 4\text{-}lept}_{\rm EFT} \, .
\label{eq:def_4fermion}
\eeq
All the operators appearing in the 4-quark, semi-leptonic and 4-lepton sectors, together with the expressions of the 
corresponding Wilson coefficients as a function of the PQ-2HDM parameters are listed in \app{sec:4fermion}.

%%%%%%%%%%%%%%%%%%%%%%%%%%%%%%%%%%%%%%%%%%%%%%%%%%%%%%%%%%%%%%%%%%%%%%%%%%%%%%%%%%%%%%%%%%%%%%%%%
\subsection{Connection between IR/UV sources of flavour violation}
\label{sec:IRUVconnection}
%%%%%%%%%%%%%%%%%%%%%%%%%%%%%%%%%%%%%%%%%%%%%%%%%%%%%%%%%%%%%%%%%%%%%%%%%%%%%%%%%%%%%%%%%%%%%%%%%

The IR sources of flavour violation, $C^{V,A}_{u,d,e}$, involving the light axion field and the UV sources, 
$\epsilon^{u,d,e}$, involving the heavy radial modes of the PQ-2HDM, turn out to be related thanks to 
\eqs{eq:chargesu}{eq:chargese}, which encode the action of the $\U(1)_{\rm PQ}$ symmetry on the Yukawa 
sector of the PQ-2HDM. By means of the SM fermion masses in \eq{eq:defMYuk} it is possible to rewrite 
the Yukawa matrices in terms of the (non-universal) fermionic PQ charges as 
\begin{align}
Y_1^u &= \frac{\sqrt{2}}{v c_\beta} \( - \X_q M_u + M_u \X_u + \X_2 M_u \) \, , \quad Y_2^u = \frac{\sqrt{2}}{v s_\beta} 
\( \X_q M_u - M_u \X_u - \X_1 M_u \) \, , \\  
Y_1^d &= \frac{\sqrt{2}}{v c_\beta} \( \X_q M_d - M_d \X_d + \X_2 M_d \) \, , \quad \ \ \ Y_2^d = \frac{\sqrt{2}}{v s_\beta} 
\( - \X_q M_d + M_d \X_d - \X_1 M_d \) \, , \\
Y_1^e &= \frac{\sqrt{2}}{v c_\beta} \( \X_\ell M_e - M_e \X_e + \X_2 M_e \) \, , \quad \ \ \ Y_2^e = \frac{\sqrt{2}}{v s_\beta} 
\( - \X_\ell M_e + M_e \X_e - \X_1 M_e \) \, , 
\end{align}
where the PQ charge relation $\X_2-\X_1=1$ arising from the scalar potential has also been used. Hence, by exploiting 
the definitions of $C^{L,R}_{f}$ and $\epsilon^{f}$ in \eq{eq:CRLf} and \eq{eq:EpsilonCouplings}, we can write the 
following expression relating the flavour violating couplings of the axion field and the heavy radial modes:
\begin{align}
\label{eq:epsuC}
\epsilon^u &=  \frac{2N \sqrt{2}}{v c_\beta} \(- C_u^L \hat M_u + \hat M_u C_u^R + \frac{\X_2}{2N} \hat M_d\)  \, ,\\
\label{eq:epsdC}
\epsilon^d &= \frac{2N \sqrt{2}}{v s_\beta} \(- C_d^L \hat M_d + \hat M_d C_d^R - \frac{\X_1}{2N} \hat M_d\) \, , \\
\label{eq:epseC}
\epsilon^e &= \frac{2N \sqrt{2}}{v s_\beta} \(- C_e^L \hat M_e + \hat M_e C_e^R - \frac{\X_1}{2N} \hat M_e\) \, . 
\end{align}
\eqs{eq:epsuC}{eq:epseC} provide the aforementioned connection between the IR and UV sources of flavour violation, 
and they apply to the most general class of renormalizable non-universal axion models based on two Higgs doublets.
Notice that these relations do not depend explicitly on the unknown fermion mass rotation matrices, but only 
on the physical fermion masses and fermion couplings to the axion. Moreover, notice that the terms proportional 
to the Higgs doublet PQ charges, $\X_{1,2}$, are diagonal and hence they do not intervene in flavour violating 
interactions. In addition, from the hermiticity of the $C_f^{L,R}$ matrices, it follows that transposed off-diagonal 
entries can be simply related through 
\bea
(\epsilon^f_{ji})^*=\epsilon^f_{ij}(m_{f_i} \leftrightarrow m_{f_j}) \, .
\eea
For a more transparent connection with flavour observables, it is useful to rewrite \eqs{eq:epsuC}{eq:epseC} in terms 
of vector and axial flavour violating couplings. For instance, the off-diagonal $d$-quark entries read:
\begin{align}
\label{eq:IRVcoupling}
(\epsilon_A^d)_{ij}\equiv\epsilon^d_{ij}+\epsilon^{d\,*}_{ji}=\sqrt{2}\,\frac{2N}{v\,s_\beta}(m_{d_i}+m_{d_j})(C^A_d)_{ij}\,, 
\qquad (i\neq j) \\
\label{eq:IRAcoupling}
(\epsilon_V^d)_{ij}\equiv\epsilon^d_{ij}-\epsilon^{d\,*}_{ji} =\sqrt{2}\,\frac{2N}{v\,s_\beta}(m_{d_i}-m_{d_j})(C^V_d)_{ij}\,. 
\qquad (i\neq j)
\end{align}
The charged-lepton coefficients are simply obtained by replacing $d \to e$ in \eqs{eq:IRVcoupling}{eq:IRAcoupling}, 
while the up-quark ones can be obtained via the substitutions $d \to u$ and $s_\beta \to c_\beta$. In addition, 
for describing $\Delta F = 2$ observables (cf.~\eqs{eq:boundK}{eq:C4d}) the following combinations are going to be useful:
\begin{align}
	\label{eq:epsilonsum} 
	(\epsilon^d_{ij})^2+(\epsilon^d_{ij})^{*\,2}&=\frac{(\epsilon_A^d)^2_{ij}+(\epsilon_V^d)^2_{ij}}{2} \nonumber \\
	&=\(\frac{2N}{vs_\beta}\)^2\[(m_{d_i}+m_{d_j})^2 (C^A_d)^2_{ij} +(m_{d_i}-m_{d_j})^2 (C^V_d)^2_{ij} \] \, , \\
	\label{eq:epsilonproduct} 
	2\epsilon^d_{ij}(\epsilon^d_{ij})^{*}&=\frac{(\epsilon_A^d)^2_{ij}-(\epsilon_V^d)^2_{ij}}{2} \nonumber \\
	&=\(\frac{2N}{vs_\beta}\)^2\[(m_{d_i}+m_{d_j})^2 (C^A_d)^2_{ij} -(m_{d_i}-m_{d_j})^2 (C^V_d)^2_{ij} \] \, . 
\end{align}
Again, the analogous expressions for charged leptons are obtained by replacing $d \to e$, while for the up-quark sector 
by sending $d \to u$ and $s_\beta \to c_\beta$ in \eqs{eq:epsilonsum}{eq:epsilonproduct}.

%%%%%%%%%%%%%%%%%%%%%%%%%%%%%%%%%%%%%%%%%%%%%%%%%%%%%%%%%%%%%%%%%%%%%%%%%%%%%%%%%%%%%%%%%%%%%%%%%
\subsection{Non-universal PQ-2HDM Yukawa textures}
\label{sec:textures}
%%%%%%%%%%%%%%%%%%%%%%%%%%%%%%%%%%%%%%%%%%%%%%%%%%%%%%%%%%%%%%%%%%%%%%%%%%%%%%%%%%%%%%%%%%%%%%%%%

The relations derived in the previous section hold for a general non-universal PQ-2HDM. On the other hand, it is 
useful to provide for the following phenomenological analysis a couple of benchmark Yukawa textures that can nicely exemplify the IR/UV flavour correlation. Here, we will 
consider two explicit examples of 2+1-type models, i.e.~with at least two generations featuring the same PQ charges, 
originally presented in Ref.~\cite{DiLuzio:2017ogq} (see also \cite{DiLuzio:2021ysg}) in order to suppress 
the axion couplings to nucleons. Despite the reasons behind their original motivation, these two textures, labelled 
respectively as M1 and M4 in Ref.~\cite{DiLuzio:2017ogq}, are prototypes of models with quark flavour violation in 
axion couplings, respectively in the left-handed (LH) and in the right-handed (RH) down-quark sector. The lepton 
sector is more model-dependent and we report here for definiteness just one of the possible choices. Note that in 
these models it is also possible to achieve electrophobia via a tuning involving the rotation matrices of the charged 
leptons \cite{DiLuzio:2017ogq}, thus realizing the so-called ``astrobhobic'' axion scenario in which both coupling 
to nucleons and electrons are suppressed, and astrophysical axion bounds are in turn suppressed. 
The M1 and M4 textures are defined by the following Yukawa matrices:  
\begin{itemize}
\item M1 model
\begin{align}
Y^u_1 &= 
\begin{pmatrix}
y^u_{11} & y^u_{12} & y^u_{13} \\
y^u_{21} & y^u_{22} & y^u_{23} \\
0 & 0 & 0 
\end{pmatrix} \, , \quad 
Y^u_2 = 
\begin{pmatrix}
0 & 0 & 0 \\
0 & 0 & 0 \\
y^u_{31} & y^u_{32} & y^u_{33} 
\end{pmatrix} \, , \\ 
Y^d_1 &= 
\begin{pmatrix}
0 & 0 & 0 \\
0 & 0 & 0 \\
y^d_{31} & y^d_{32} & y^d_{33} 
\end{pmatrix} \, , \quad 
Y^d_2 = 
\begin{pmatrix}
y^d_{11} & y^d_{12} & y^d_{13} \\
y^d_{21} & y^d_{22} & y^d_{23} \\
0 & 0 & 0 
\end{pmatrix} \, , \\ 
Y^e_1 &= 
\begin{pmatrix}
y^e_{11} & y^e_{12} & y^e_{13} \\
y^e_{21} & y^e_{22} & y^e_{23} \\
0 & 0 & 0 
\end{pmatrix} \, , \quad 
Y^e_2 = 
\begin{pmatrix}
0 & 0 & 0 \\
0 & 0 & 0 \\
y^e_{31} & y^e_{32} & y^e_{33} 
\end{pmatrix} \, .
\end{align}
Upon a baryon and lepton number transformation one can set $\X_{q_\alpha} = \X_{\ell_\alpha} = 0$ ($\alpha = 1,2$), 
so that the PQ charge matrices satisfying the constraints in \eqs{eq:chargesu}{eq:chargese} read
\begin{align}
\label{eq:PQchargesM1}
\X_q & = \diag (0,0,1)  \, ,  \quad  \X_u = \diag (s^2_\beta,s^2_\beta,s^2_\beta)   \, , \quad 
\X_d = \diag (c^2_\beta,c^2_\beta,c^2_\beta) \, , \nonumber \\  
\X_\ell &= - \X_q \, ,  \quad \quad \quad \quad \X_e = - \X_u \, . 
\end{align}
which shows that in the M1 model only the PQ charges in the quark and lepton LH sectors are non-universal. 

\item M4 model 
\begin{align}
Y^u_1 &= 
\begin{pmatrix}
y^u_{11} & y^u_{12} & y^u_{13} \\
y^u_{21} & y^u_{22} & y^u_{23} \\
y^u_{31} & y^u_{32} & y^u_{33} 
\end{pmatrix} \, , \quad 
Y^u_2 = 0 \, , \\ 
Y^d_1 &= 
\begin{pmatrix}
0 & y^d_{12} & y^d_{13}\\
0 & y^d_{22} & y^d_{23} \\
0 & y^d_{32} & y^d_{33}
\end{pmatrix} \, , \quad 
Y^d_2 = 
\begin{pmatrix}
y^d_{11} & 0 & 0\\
y^d_{21} & 0 & 0 \\
y^d_{31} & 0 & 0
\end{pmatrix}  \, , \\ 
Y^e_1 &= 
\begin{pmatrix}
y^e_{11} & y^e_{12} & y^e_{13} \\
0 & 0 & 0 \\
0 & 0 & 0 
\end{pmatrix} \, , \quad 
Y^e_2 = 
\begin{pmatrix}
0 & 0 & 0 \\
y^e_{21} & y^e_{22} & y^e_{23} \\
y^e_{31} & y^e_{32} & y^e_{33} 
\end{pmatrix}
 \, .
\end{align}
Setting $\X_{q_1} = \X_{\ell_1} = 0$ via a baryon and lepton number transformation, the PQ charges in the M4 model 
are fixed to be 
\begin{align}
\label{eq:PQchargesM4}
\X_q &= \diag (0,0,0) \, ,  \quad \quad
\X_u = \diag (s^2_\beta,\,s^2_\beta,\,s^2_\beta) \, , \quad
\X_d = \diag (c^2_\beta,\,-s^2_\beta,\,-s^2_\beta) \, ,  \nonumber \\
\X_\ell &= - \diag (0,1,1) \, ,  \quad
\X_e = - \X_u \, , 
\end{align}
which imply non-universality in the RH down sector and in the LH lepton sector. 
\end{itemize}

\noindent 
With the specific choice of the trilinear $H_2^\dag H_1 \phi$ term in the scalar potential, one has $2N=1$ both 
in the M1 and M4 models. Flavour violating axion couplings to fermions (as defined in \eq{eq:CRLf}) arise 
exclusively in the LH sector for the M1 model, i.e. 
\begin{align}
(C^{V,A}_u)_{i \neq j} &= \pm (V_{u_L}^\dag \X_{q} V_{u_L})_{i \neq j} \, , \\ 
(C^{V,A}_d)_{i \neq j} &= \pm (V_{d_L}^\dag \X_{q} V_{d_L})_{i \neq j} \, ,  \\ 
(C^{V,A}_e)_{i \neq j} &= \pm (V_{e_L}^\dag \X_{\ell} V_{e_L})_{i \neq j} \, , 
\end{align}
while for the M4 model flavour violation arises in the RH down-quark sector and LH charged-lepton sector respectively: 
\begin{align}
(C^{V,A}_u)_{i \neq j} &= 0 \, , \\ 
(C^{V,A}_d)_{i \neq j} &= (V_{d_R}^\dag \X_{d} V_{d_R})_{i \neq j} \, , \\
(C^{V,A}_e)_{i \neq j} &= \pm (V_{e_L}^\dag \X_{\ell} V_{e_L})_{i \neq j} \, . 
\end{align}
Moreover, the combination of flavour violating couplings appearing in $\Delta F=2$ observables, defined in 
\eqs{eq:epsilonsum}{eq:epsilonproduct}, 
take the simpler form (both in the M1 and M4 models)
\begin{align}
\label{eq:M13Sum} 
(\epsilon^d_{ij})^2+(\epsilon^d_{ij})^{*\,2}&=2\(\frac{1}{vs_\beta}\)^2 (m_{d_i}^2+m_{d_j}^2) (C^{V,A}_d)^2_{ij} \, , \\
\label{eq:M13Product} 
2\,\epsilon^d_{ij}(\epsilon^d_{ij})^{*}&=4\(\frac{1}{vs_\beta}\)^2m_{d_i} \,m_{d_j} (C^{V,A}_d)^2_{ij} \, ,
\end{align}
with the analogous expressions for the up-quark sector obtained by replacing $d \to u$ and $s_\beta \to c_\beta$.

%%%%%%%%%%%%%%%%%%%%%%%%%%%%%%%%%%%%%%%%%%%%%%%%%%%%%%%%%%%%%%%%%%%%%%%%%%%%%%%%%%%%%%%%%%%%%%%%%%%%%%%%%%%%%%%%%%%
%%%%%%%%%%%%%%%%%%%%%%%%%%%%%%%%%%%%%%%%%%%%%%%%%%%%%%%%%%%%%%%%%%%%%%%%%%%%%%%%%%%%%%%%%%%%%%%%%%%%%%%%%%%%%%%%%%%

\section{Flavour phenomenology}
\label{sec:pheno}

%%%%%%%%%%%%%%%%%%%%%%%%%%%%%%%%%%%%%%%%%%%%%%%%%%%%%%%%%%%%%%%%%%%%%%%%%%%%%%%%%%%%%%%%%%%%%%%%%%%%%%%%%%%%%%%%%%%
%%%%%%%%%%%%%%%%%%%%%%%%%%%%%%%%%%%%%%%%%%%%%%%%%%%%%%%%%%%%%%%%%%%%%%%%%%%%%%%%%%%%%%%%%%%%%%%%%%%%%%%%%%%%%%%%%%%

With the aid of the relations obtained in the previous section we can now investigate the interplay between axion flavour 
phenomenology and the signatures arising from the heavy Higgs radial modes of the PQ-2HDM. We start by discussing 
LHC constraints on low-scale, i.e.~$\mathcal{O}(1\, \text{TeV})$, 2HDM and then consider all relevant flavour bounds 
on the non-universal PQ-2HDM, both in the quark and lepton sectors. This phenomenological analysis actually applies 
to a general 2HDM with tree-level flavour changing couplings, regardless of the presence of the PQ symmetry. 
All the details regarding the 2HDM potential and its mass spectrum are deferred to \app{sec:scalpot}, while the 
analysis of meson mixing and a collection of formulae for LFV processes can be found in \app{sec:mesonmixing} and \ref{app:LFVobs}. Finally, based on the IR/UV flavour connection (cf.~\eqs{eq:epsuC}{eq:epseC}), we discuss the 
consequences of an axion discovery in flavour violating decays on the PQ-2HDM parameter space. 

%%%%%%%%%%%%%%%%%%%%%%%%%%%%%%%%%%%%%%%%%%%%%%%%%%%%%%%%%%%%%%%%%%%%%%%%%%%%%%%%%%%%%%%%%%%%%%%%%%%%%%%%%%%%%%%%%%%
\subsection{LHC bounds on the 2HDM}
%%%%%%%%%%%%%%%%%%%%%%%%%%%%%%%%%%%%%%%%%%%%%%%%%%%%%%%%%%%%%%%%%%%%%%%%%%%%%%%%%%%%%%%%%%%%%%%%%%%%%%%%%%%%%%%%%%%

The 2HDM scalar sector can be tested at the LHC in two different ways: either indirectly via the modification of Higgs 
couplings or by directly hunting for new scalar resonances. The first type of searches are only sensitive to the 
mixing angles in the scalar sector, while direct searches are also sensitive to the typical mass scale of the 
enlarged scalar sector. 

LHC Higgs data implies that the observed properties of the $h(125)$ resonance are very close to those predicted by the SM. 
In particular, observables such as Higgs couplings to fermions and SM gauge bosons yield already a robust upper bound 
on the rotation matrix that diagonalize the neutral scalar sector, $c_{\alpha-\beta} \lesssim 0.3(0.1)$ \cite{ATLAS:2021vrm} 
for a type-I(II) 2HDM  \cite{Hall:1981bc}.\footnote{Recall that (universal) 2HDM's with natural flavour conservation are 
conventionally classified as $i)$ type-I model if only one of the two scalar doublets couples (universally) to all fermions 
or $ii)$ type-II model if up- and down-quarks sectors couple to different doublets. In particular, the (universal) DFSZ 
axion model is compatible only with the type-II structure.} 

In order to automatically satisfy LHC constraints on Higgs couplings, the so-called SM \textit{alignment} limit \cite{Gunion:2002zf} 
has been introduced, in which one of the two Higgs doublets is nearly aligned with the electroweak VEV, thus reproducing 
the properties of a SM-like Higgs. There are basically two ways of achieving alignment. The first one is \textit{alignment 
with decoupling}, in which the condition $c_{\alpha-\beta}\to 0$ is obtained by decoupling all the heavy radial modes, 
$m_H \simeq m_A \simeq m_{H^+} \gg m_h$ (cf.~\eq{eq:massrelations3}). In this case we identify $h$ and $H$ respectively 
as the SM-like and non-SM-like Higgs scalars. \textit{Alignment without decoupling} is obtained, instead, by taking 
the scalar potential parameter $\Lambda_6\to 0$ which allows for a second light Higgs in the spectrum with a mass 
either larger \cite{Bernon:2015qea} (if $c_{\alpha-\beta}\to 0$) or smaller \cite{Bernon:2015wef} (if $s_{\alpha-\beta} 
\to 0$) than the observed scalar resonance at 125 GeV. The field $H$ then labels the non-SM-Higgs (SM-Higgs) in the 
first (latter) eventuality. Notice that in the alignment without decoupling scenario the mass spectrum of the scalar 
particles is not necessarily degenerate and it is determined, in particular, by the parameter $\Lambda_5$ of the scalar 
potential. Hereafter, we will only consider the case $m_H\gtrsim m_h$ (i.e. $c_{\alpha-\beta}\simeq 0$) and therefore 
$h$ will always denote the SM-like Higgs. 

LHC has also greatly contributed to direct searches for an extended scalar sector, looking for both neutral~\cite{ATLAS:2017eiz,
ATLAS:2017jag,ATLAS:2017otj,ATLAS:2017tlw,ATLAS:2017uhp,ATLAS:2017xel,ATLAS:2018oht,ATLAS:2020azv,ATLAS:2020tlo,ATLAS:2020zms,
CMS:2017aza,CMS:2017hea,CMS:2018ipl,CMS:2019bnu,CMS:2019kca,CMS:2019ogx,CMS:2019pzc,CMS:2019qcx,CMS:2020jeo} and 
charged scalar particles~\cite{ATLAS:2020zzb,ATLAS:2021upq,CMS:2019bfg,CMS:2019lwf}. Different mechanisms can be responsible 
for the production of new scalar resonances: $i)$ Vector Boson Fusion (VBF) is proportional to $c_{\alpha-\beta}$ and hence 
substantially suppressed in the alignment limit, $ii)$ gluon-gluon Fusion (ggF) depends, instead, parametrically on $t_\beta$  
and therefore bottom quarks can contribute sizeably to ggF. 
Direct and indirect searches, as well as different production mechanisms are hence used to give bounds on the 2HDM 
in the $(m_X, t_\beta)$ plane (where $X = H, A, H^+$), for a given value of $c_{\alpha-\beta}$. For instance, searches 
for scalar particle decays into gauge bosons test the lower parts of the $(m_X, t_\beta)$ plane for type-I and II 
models~\cite{ATLAS:2017jag,ATLAS:2017uhp,CMS:2019bnu,ATLAS:2017tlw,ATLAS:2017otj,ATLAS:2020tlo}, while type-II enhanced 
di-$\tau$ decay set a strong upper bound on $t_\beta$ \cite{ATLAS:2017eiz,ATLAS:2020zms}. For recent analyses on the 
parameter space of the 2HDM, see e.g.~\cite{Wang:2022yhm,Atkinson:2021eox,Atkinson:2022pcn}. In particular, a recent 
global fit to type-II 2HDM \cite{Atkinson:2022pcn} shows that $m_H\gtrsim 500\,$GeV independently of $t_\beta$, 
while the bound stretches up to $m_H\simeq 1\,$TeV for $t_\beta \gtrsim 1$.

One can partially use type-I and II 2HDM collider searches to provide approximate bounds for the 
non-universal PQ-2HDM models by applying the present limits on the 33 entries of the Yukawa matrices ($y_{t,b,\tau}$)
of the benchmark M1 and M4 models introduced in \sect{sec:nonuniversalaxion}. 
In this approximation the type-I and II models searches can be related to the M1 and M4 Yukawa structures, as summarized 
in \Table{tab:comparisonYukCoupl} for the 3rd generation fermion couplings to the $H$ scalar field. However, although M1 and 
M4 models behave only partially as type-I and II 2HDM, in the following, as a conservative approach, we will regardless  employ bounds from 
the type-II model, which turn out to be more stringent. 

\begin{table}[h]
\centering
\begin{tabular}{c | c | c | c}
 Model & $y_t$ & $y_b$ & $y_\tau$\\ 
 \hline  
 \rule{0pt}{3ex}
 \cellcolor{blue!25}Type-I & $\dfrac{s_\alpha}{s_\beta}$ & $\dfrac{s_\alpha}{s_\beta}$ & $\dfrac{s_\alpha}{s_\beta}$  \\[0.25cm]
 \rule{0pt}{3ex}
 \cellcolor{red!25}Type-II & $\dfrac{s_\alpha}{s_\beta}$ & $\dfrac{c_\alpha}{c_\beta}$ &  $\dfrac{c_\alpha}{c_\beta}$ \\[0.25cm]
 \hline
 \hline
 \rule{0pt}{3ex}
 M1 &  \cellcolor{nicepurple!40}$\dfrac{s_\alpha}{s_\beta}$ &  \cellcolor{red!25} $\dfrac{c_\alpha}{c_\beta}$ &   \cellcolor{blue!25}$\dfrac{s_\alpha}{s_\beta}$ \\[0.25cm]
 \rule{0pt}{3ex}
 M4 &  \cellcolor{gray!50}$\dfrac{c_\alpha}{c_\beta}$ & \cellcolor{blue!25} $\dfrac{s_\alpha}{s_\beta}$ &  \cellcolor{red!25} $\dfrac{c_\alpha}{c_\beta}$   \\
 \end{tabular}
\caption{Scaling of the $H$ couplings to 3rd generation SM fermions, in the type-I and II 2HDM compared to the M1 and M4 models.
Type-I (II) couplings are emphasized in blue (red), while grey shaded couplings have no counterpart in universal 2HDMs. Couplings which are the same in both models are colored in purple.}
 \label{tab:comparisonYukCoupl}
\end{table}
%%%%%%%%%%%%%%%%%%%%%%%%%%%%%%%%%%%%%%%%%%%%%%%%%%%%%%%%%%%%%%%%%%%%%%%%%%%%%%%%%%%%%%%%%%%%%%%%%%%%%%%%%%%%%%%%%%%
\subsection{Flavour constraints on the non-universal 2HDM} 
%%%%%%%%%%%%%%%%%%%%%%%%%%%%%%%%%%%%%%%%%%%%%%%%%%%%%%%%%%%%%%%%%%%%%%%%%%%%%%%%%%%%%%%%%%%%%%%%%%%%%%%%%%%%%%%%%%%

The strongest constraints on the quark flavour violating couplings of the heavy radial modes of the PQ-2HDM 
arise from neutral mesons mixing. The parameter space of a flavourful 2HDM has been previously studied also in 
connection with LHC data (see for example \cite{Babu:2018uik}). Previous analyses often relied on Yukawa textures 
specifically introduced in order to suppress tree-level flavour changing neutral currents (FCNCs), as e.g.~in the 
case of the Cheng-Sher ansatz \cite{Cheng:1987rs}. In this Section, conversely, we do not assume any particular 
form for the Yukawa matrices, 
and perform, instead, an analysis of meson mixing constraints on the parameter space of a general 2HDM with 
tree-level flavour-changing couplings. Our numerical analysis, including QCD running effects, follows the one in 
Ref.~\cite{Babu:2018uik} (see also \cite{Ciuchini:1998ix,Becirevic:2001jj,UTfit:2007eik}). Additionally, we take 
into account (whenever available) the theoretical error on the SM predictions for the meson mixing parameters. 
In the case of $B$ systems theoretical uncertainties are at the level of  $\sim 5\%$ \cite{DiLuzio:2019jyq}, while 
Kaon physics has not yet reached this level of precision and lattice calculations provide $K$-form factors with an 
$\mathcal{O}(30\%)$ error, see for example  Refs.~\cite{FlavourLatticeAveragingGroupFLAG:2021npn,Brod:2011ty}.  

The details of the analysis are reported in \app{sec:mesonmixing}. Here, we summarize the main results in the 
alignment scenario, which is necessary in order to satisfy LHC constraints on $c_{\alpha - \beta}$. The bounds 
from meson mixing can be recast in this limit as

\begin{align}
    \label{eq:BoundDecouplingKaon2}
  \left(\frac{[4.1,\, 3.6]\,\text{TeV}}{s_{2\beta}\, m_H}\right)^2\frac{m_H^2+m_A^2}{ 2m_A^2 }\left\vert\,\frac{\epsilon_{12}^d \epsilon^{d\,*}_{21}}{y_s^2\,\lambda^{2}}+0.12\frac{m_H^2-m_A^2}{m_H^2+m_A^2}\frac{(\epsilon_{12}^d)^2+ (\epsilon^{d\,*}_{21})^2}{y_s^2\,\lambda^{2}}\right\vert< 1 \, , \\
    \label{eq:BoundDecouplingBd2}
       \left(\frac{[1.4,\, 1.7]\,\text{TeV}}{s_{2\beta}\, m_H}\right)^2\frac{m_H^2+m_A^2}{2m_A^2 }\left\vert\,\frac{\epsilon_{13}^d \epsilon^{d\,*}_{31}}{y_b^2\lambda^{6}}+0.17\frac{m_H^2-m_A^2}{m_H^2+m_A^2}\frac{(\epsilon_{13}^d)^2+ (\epsilon^{d\,*}_{31})^2}{y_b^2\lambda^{6}}\right\vert <1 \, ,\\
    \label{eq:BoundDecouplingBs2}
     \left(\frac{[1.5,\, 1.6]\,\text{TeV}}{s_{2\beta}\, m_H}\right)^2\frac{m_H^2+m_A^2}{ 2 m_A^2 }\left\vert\,\frac{\epsilon_{23}^d \epsilon^{d\,*}_{32}}{y_b^2 \lambda^{4}}+0.17\frac{m_H^2-m_A^2}{m_H^2+m_A^2}\frac{(\epsilon_{23}^d)^2+ (\epsilon^{d\,*}_{32})^2}{y_b^2 \lambda^{4}}\right\vert< 1 \, ,\\
    \label{eq:BoundDecouplingD2}
      \left(\frac{23 \,\text{TeV}}{m_H}\right)^2\frac{m_H^2+m_A^2}{ 2m_A^2 }\left\vert\,\frac{\epsilon_{12}^u \epsilon^{u\,*}_{21}}{ y_c^2\lambda^{2}}+0.15\frac{m_H^2-m_A^2}{m_H^2+m_A^2}\frac{(\epsilon_{12}^u)^2 +(\epsilon^{u\,*}_{21})^2}{y_c^2\lambda^{2}}\right\vert< 1 \, ,
\end{align}
where we have equated the contribution of the Wilson coefficients of the releavant 4-fermion operators to the mass scale 
written in the bracket and rescaled the UV sources of flavour violation $(\epsilon^{u,d})$ with suitable powers of Yukawas, 
$y_{d_i(u_i)}=\frac{\sqrt{2} m_i}{v s_\beta (c_\beta)}$, and Cabibbo angle, $\lambda$. This rescaling corresponds to a CKM-like 
structure for the LH rotation matrices, $V_{u_L} \sim V_{d_L} \sim V_{\rm CKM}$, and $V_{u_R} \sim V_{d_R} \sim \mathbbm{1}$ 
for the RH ones. For example, in \eq{eq:BoundDecouplingD2} one has $\epsilon_{12}^u=(V_{u_L}^\dagger Y_1^u V_{u_R})_{12}\sim 
(V_{u_L})_{12}^* (Y^u_1)_{22} (V_{u_R})_{22} \sim \lambda \frac{\sqrt{2}\, m_c}{v\,c_\beta}$. To obtain the numerical values 
reported in \eqs{eq:BoundDecouplingKaon2}{eq:BoundDecouplingD2} we have used the PDG values for $m_i(\mu)$, with $\mu =
\{2,\, 2.8,\, 4.6\}$ GeV respectively for $K$, $D$ and $B$ meson oscillations and $\lambda=0.2$. For down-type 
flavour transitions, where reliable SM theoretical predictions are available, the number quoted in the left (right) 
of the square bracket corresponds to a negative (positive) contribution from new physics. For \Dmix oscillation 
we have required, instead, that the new physics contribution would not overshoot the available experimental bound. 

Note that one of the mass terms (e.g.~$m_A$) can be traded, via \eq{eq:Lambda5}, for the scalar potential parameter 
$\Lambda_5$, subject to the perturbativity constraint $\Lambda_5 \lesssim 4\pi$. The alignment scenario with decoupling 
is readily recovered by taking the $m_H^2 \to m^2_A$ limit in \eqs{eq:BoundDecouplingKaon2}{eq:BoundDecouplingD2}. 

Employing a CKM-like structure for the fermion rotation matrices, we have studied the implication of the meson mixing 
bounds on the benchmark models M1 and M4 (see \app{app:M1M4bounds} for details). In summary, the CKM suppression is sufficient 
in order to make these models accessible at the LHC. For M1 the bound from meson mixing is $m_H\gtrsim 200$ GeV, while for 
M4 is $m_H\gtrsim 1$ TeV. This shows that, despite having FCNCs at tree level, certain PQ charge arrangements and flavour 
ansatz can relax the bounds on the mass scale of the (non-universal) 2HDM, similarly to the case of BGL models 
\cite{Branco:1996bq} (see also \cite{Branco:2011iw}). In \app{app:M1M4bounds} we have also considered the meson mixing bounds 
induced by the SM-like Higgs in the $(c_{\alpha-\beta}, t_\beta)$ plane, in the case in which the heavy Higgs ($H$) 
is not completely decoupled. These bounds are found to constrain similar regions of parameter space, as those arising 
from the SM-like Higgs LFV decays, considered in Ref.~\cite{Badziak:2021apn}.  

Regarding instead the leptonic sector of the (non-universal) 2HDM with tree-level FCNCs, we can mostly rely on previous 
analyses, such as those in Refs.~\cite{Crivellin:2013wna,Crivellin:2013hpa}. Indeed, the bounds from $l_i\to l_j \gamma$ 
\cite{BaBar:2009hkt,MEG:2013oxv} have been only marginally improved \cite{MEG:2016leq,Belle:2021ysv}; while leptonic 
decays of the type $l_i\to 3l_j$ or $l_i\to l_j l_k l_k$ are still based on the SINDRUM \cite{SINDRUM:1987nra} and 
Belle \cite{Hayasaka:2010np} experiments, also taken into account in \cite{Crivellin:2013hpa}. On the other hand, the 
limits on Higgs LFV decays have been considerably improved in the recent years by the LHC \cite{CMS:2021rsq,ATLAS:2023mvd}. 
In particular, very strong limits on the parameter space of the (non-universal) 2HDM arise from the $h \to \tau \mu$ 
and $h \to \tau e$ processes, while for the case of $\mu$-$e$ transitions $\mu \to e \gamma$ still provides the 
most relevant constraint. The Higgs LFV branching ratios, in the notation of \eq{eq:LyukM1v2}, read
\beq
\label{eq:BRhiggsGeneral}
	\text{BR}(h\to l_i l_j)=\frac{m_h}{16\pi\Gamma_h}(|C^{h_e}_{ij}|^2+|C^{h_e}_{ji}|^2) \, , 
\eeq
where $\Gamma_h$ could receive corrections from new physics (cf.~\eq{eq:ModHiggsWidth}). By means of \eq{eq:wilson_coef2} 
we can rewrite the $2\sigma$ bound from $h \to \tau \mu/e$ directly on the $\epsilon^e$ coefficients as
\beq
     \frac{4.1\,\text{MeV}}{\Gamma_h}\frac{|\epsilon^e_{i3}|^2+|\epsilon^e_{3i}|^2}{y_\tau^2}\(\frac{c_{\alpha-\beta}}{s_\beta c_\beta}\)^2 \lesssim 0.066 \, , 
\eeq
where $i=1,\,2$ (since at $2\sigma$ both the $\mu$ and $e$ channels yield a similar bound) and 
$y_\tau= \sqrt{2} m_\tau /v s_\beta$. The Higgs branching ratio has been normalized to the SM model one, 
$\Gamma_h^{\rm SM} \simeq 4.1\,$MeV.

%%%%%%%%%%%%%%%%%%%%%%%%%%%%%%%%%%%%%%%%%%%%%%%%%%%%%%%%%%%%%%%%%%%%%%%%%%%%%%%%%%%%%%%%%%%%%%%%%%%%%%%%%%%%%%%%%%%
\subsection{IR/UV quark flavour connection} \label{sec:IRUV_meson}
%%%%%%%%%%%%%%%%%%%%%%%%%%%%%%%%%%%%%%%%%%%%%%%%%%%%%%%%%%%%%%%%%%%%%%%%%%%%%%%%%%%%%%%%%%%%%%%%%%%%%%%%%%%%%%%%%%%

As seen in \sect{sec:IRUVconnection} there is a direct relation between the IR sources of flavour violation  
($C_{f}^{V,A}$) and the UV ones ($\epsilon^{f}_{V,A}$). 
This is particularly useful in the case of a positive detection of a new physics effect, since once the strength 
of the flavour violation is known in one of the two sectors, it can be directly translated into the other one. 
The second important point is the generality of these correlations. While the flavour violating axion signals 
depend on the specific model via the PQ charges and the fermion mixing matrices $V_{f_{L/R}}$, the correlations 
derived in \sect{sec:IRUVconnection} are model-independent, in the sense that they directly relate the flavour 
coefficients entering the physical observables. 

In the following, we highlight the IR/UV connection in a class of  
non-universal PQ models with a 2+1 structure of the PQ charges, 
like the M1 and M4 models introduced in \sect{sec:textures},
exemplifications of scenarios in which flavour violation in the 
axion couplings arises either in the LH or RH sector.
The generalization to the most general setup with both LH and RH flavour violation is then straightforward. 
Through \eqs{eq:epsilonsum}{eq:epsilonproduct} we can directly translate the information on the PQ symmetry breaking scale $f_a$, appearing explicitly in the IR axion couplings to fermions into limits
on the heavy 2HDM scale $m_H$ (and $t_\beta$) by relating flavour violating axion observables with meson mixing ones:
\begin{align}
 \label{eq:rel1}
 \(\frac{f_a}{10^{11}\, \text{GeV}}\)^2\(\frac{1\, \text{TeV}}{s_{2\beta}\, m_H}\)^2\(\frac{\text{BR}(K\to\pi a)}{7.3\cdot10^{-11}}\)=\frac{2\,|M^{\text{NP}}_{12}|}{3.5\cdot10^{-15}\, \text{GeV}} \, , \\
 \label{eq:rel2}
 \(\frac{f_a}{2.8\cdot 10^7\, \text{GeV}}\)^2\(\frac{1\, \text{TeV}}{s_{2\beta}\, m_H}\)^2\(\frac{\text{BR}(B\to\pi a)}{2.3\cdot10^{-5}}\)=\frac{2\,|M^{\text{NP}}_{12}|}{3.3\cdot10^{-13}\, \text{GeV}}  \, , \\
  \label{eq:rel3}
 \(\frac{f_a}{8.8\cdot 10^7\, \text{GeV}}\)^2\(\frac{1\, \text{TeV}}{s_{2\beta}\, m_H}\)^2\(\frac{\text{BR}(B\to K a)}{7.1\cdot10^{-6}}\)=\frac{2\, |M^{\text{NP}}_{12}|}{1.2\cdot10^{-11}\, \text{GeV}} \, , \\
 \label{eq:rel4}
 \(\frac{f_a}{1.2\cdot 10^7\, \text{GeV}}\)^2\(\frac{1\, \text{TeV}}{s_{2\beta}\, m_H}\)^2\(\frac{\text{BR}(D \to\pi a)}{8\cdot10^{-6}}\)=\frac{2\,|M^{\text{NP}}_{12}|}{6.7\cdot10^{-15}\, \text{GeV}} \, .
\end{align}
These expressions have been written, for simplicity, in the alignment limit (i.e. $c_{\alpha-\beta} \simeq 0$) assuming $m_H^2\simeq m_A^2$. Flavour violating axion observables and meson mixing 
parameters have been normalized to current experimental limits.  
In particular, the 90\% CL limits on the branching ratios involving the axion filed are taken from Ref.~\cite{MartinCamalich:2020dfe}.  

\begin{figure}[t!]
    \centering
    \includegraphics[width=0.8\textwidth]{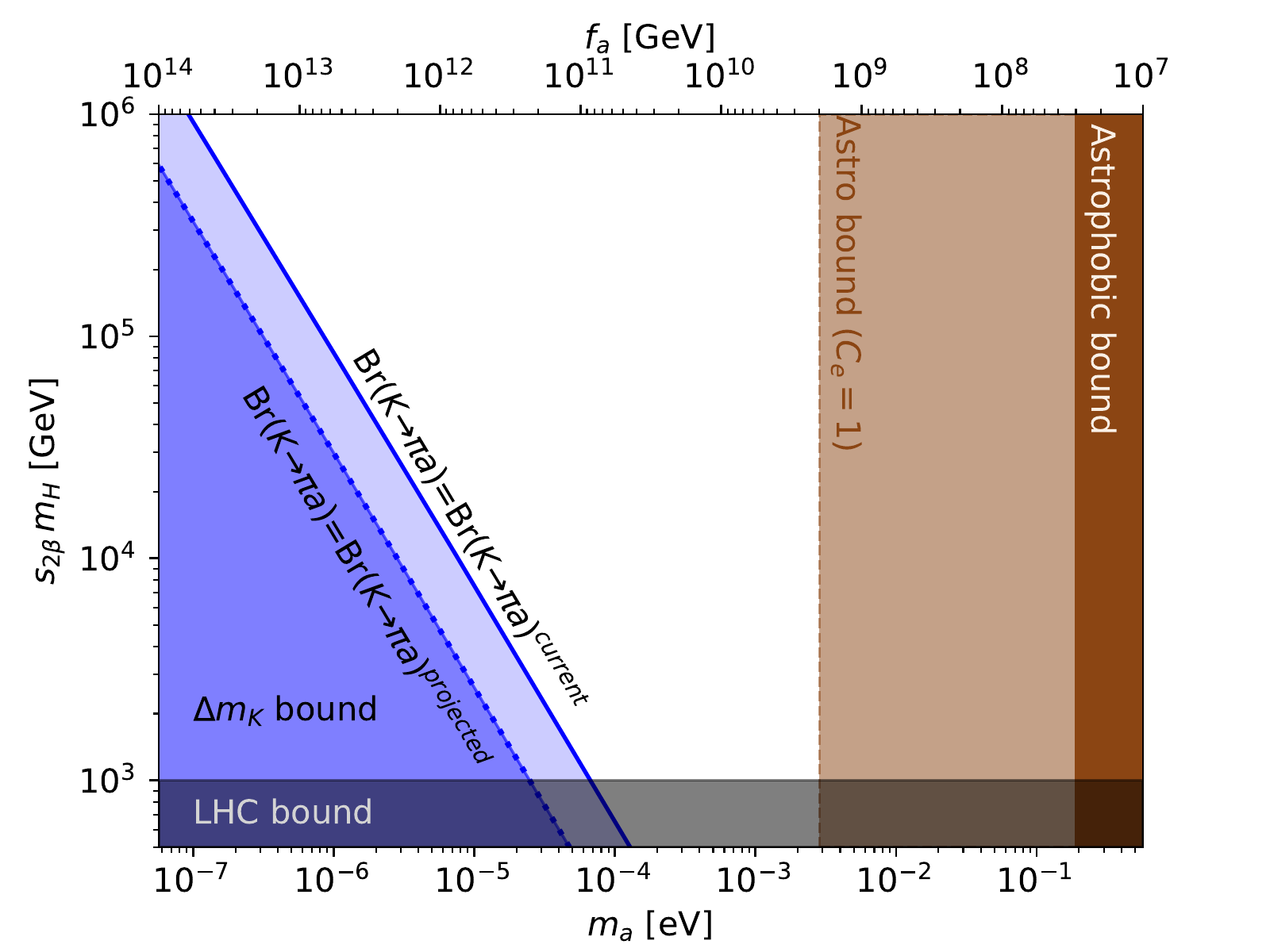}
    \\ \vspace{0.5cm}
    \includegraphics[width=0.8\textwidth]{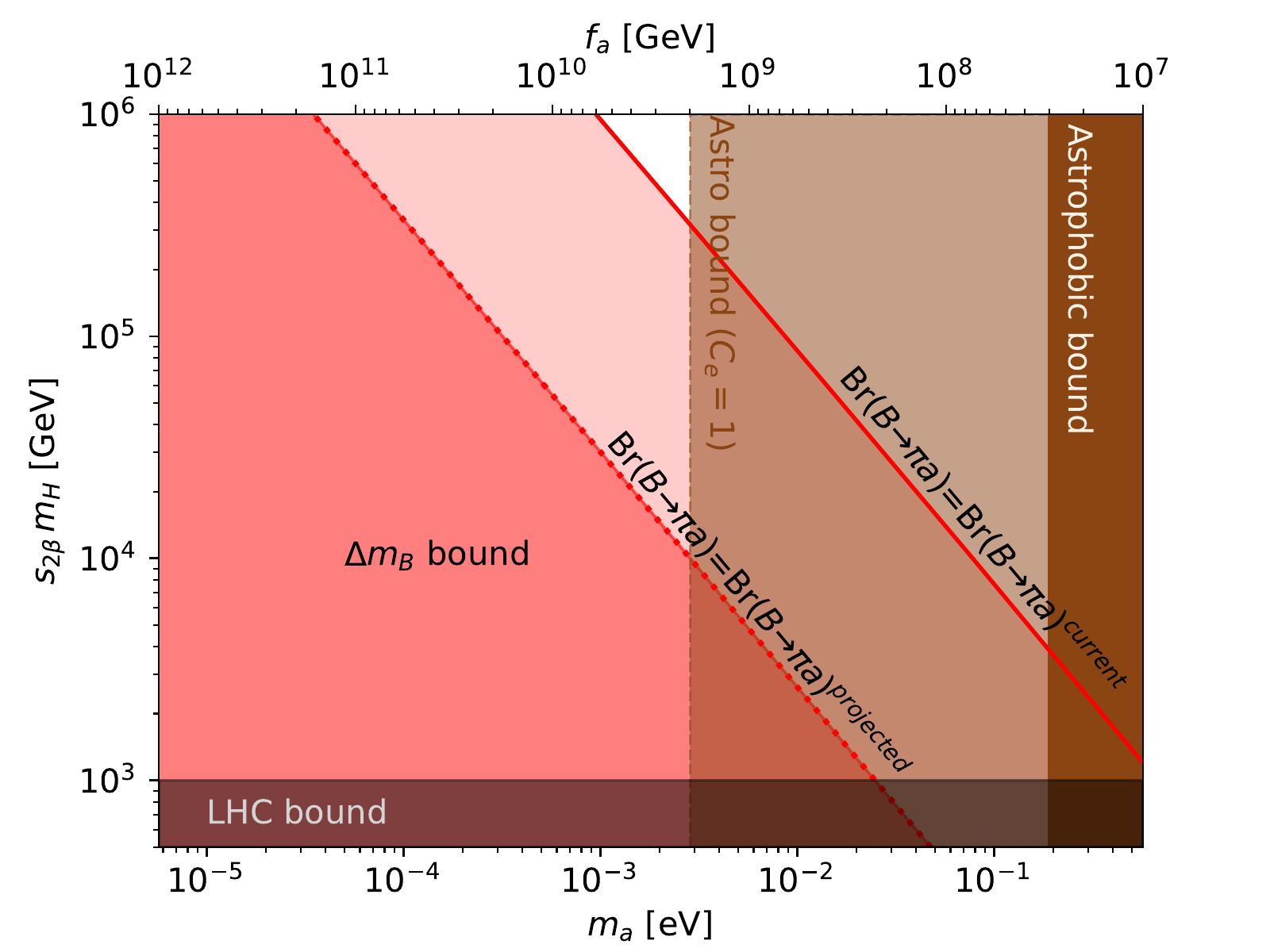}
\caption{IR/UV quark flavour connection in the $(m_a, s_{2\beta} m_H)$ plane, 
assuming an axion discovery in a given flavour transition, with solid (dotted) lines corresponding to current bounds (future sensitivities) on meson decays. 
The upper-right region is allowed by current bounds on $K^0$ (upper plot) and $B^0_d$ (lower plot) meson mixing. LHC bounds (set approximately at 1 TeV) are displayed in grey, while the brown regions correspond to astrophysical bounds on the axion decay constant, between $f_a \sim 3 \times 10^7\,$GeV (astrophobic scenario) and $f_a\sim 2\times 10^9\,$GeV, an axion with a sizeable coupling to electrons ($C_e = 1$).}
    \label{fig:interplayUVIR1}
\end{figure}%
\begin{figure}[t!]
    \centering
    \includegraphics[width=0.8\textwidth]{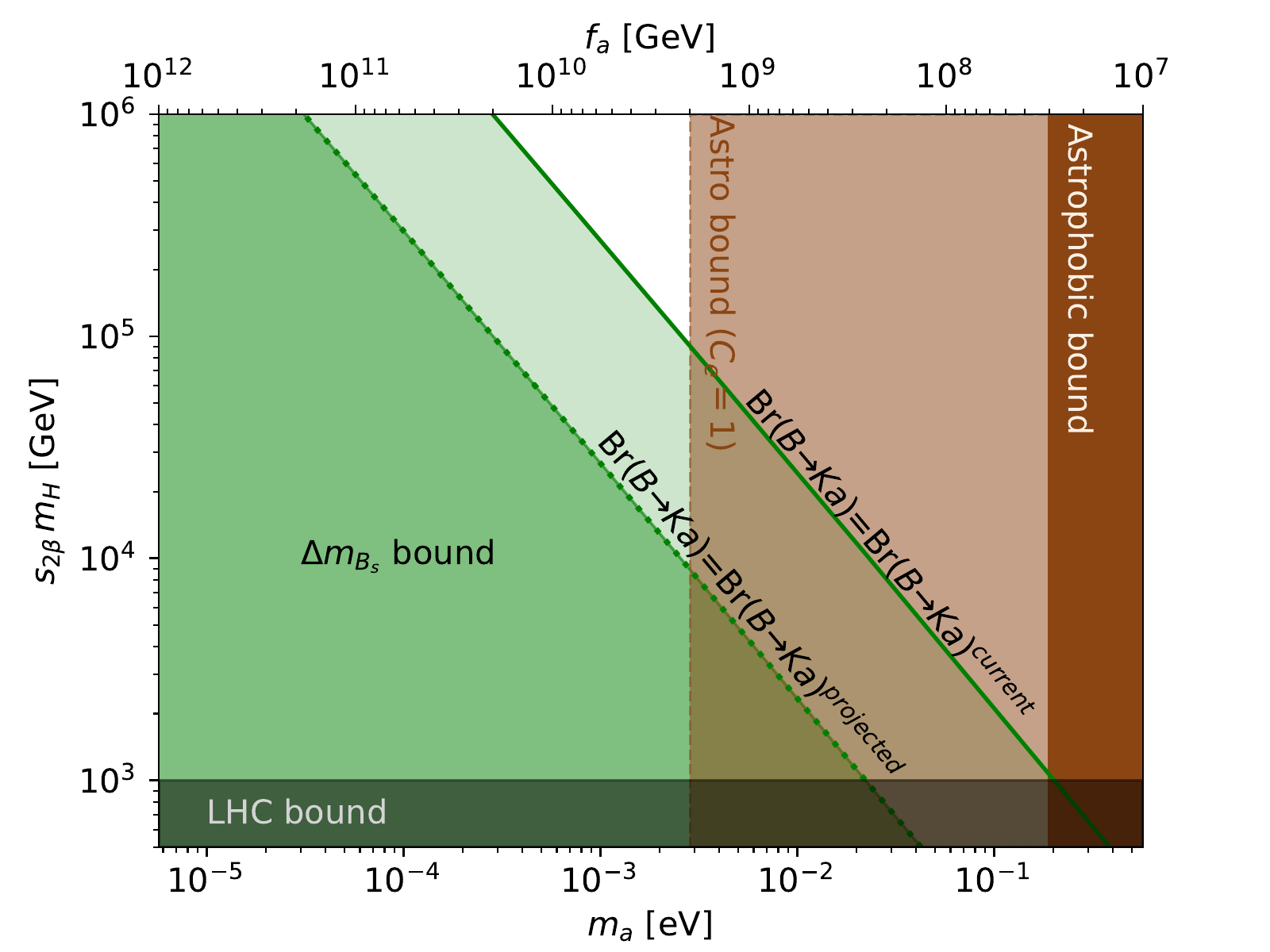} 
    \\ \vspace{0.5cm}
    \includegraphics[width=0.8\textwidth]{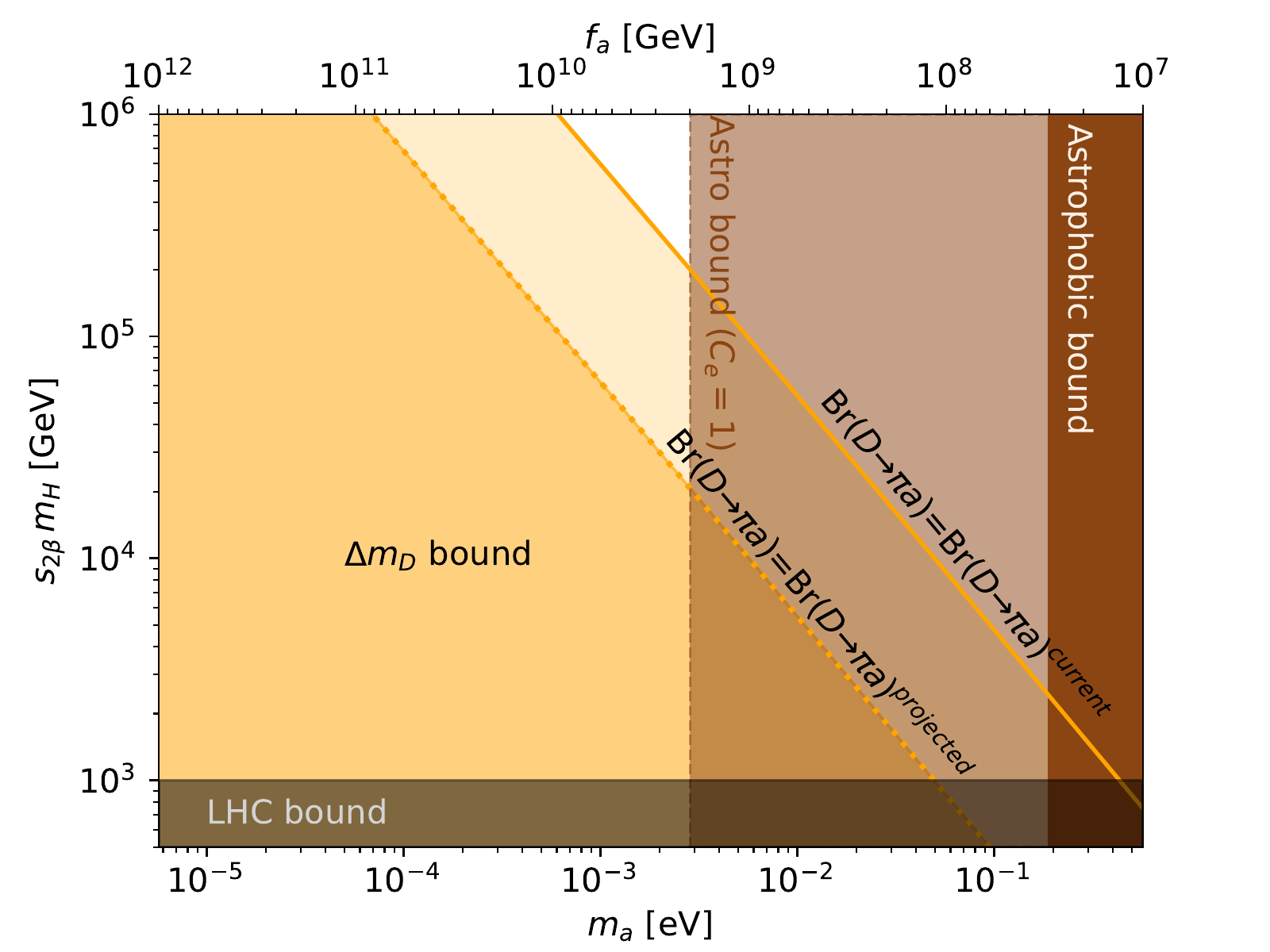}
\caption{IR/UV quark flavour connection in the $(m_a, s_{2\beta} m_H)$ plane, 
assuming an axion discovery in a given flavour transition, 
with solid (dotted) lines corresponding to current bounds (future sensitivities) on meson decays. The upper-right region is allowed by current bounds on $B^0_s$ (upper plot) and $D^0$ (lower plot) meson mixing. LHC bounds (set approximately at 1 TeV) are displayed in grey, while the brown regions correspond to astrophysical bounds on the axion decay constant, between $f_a \sim 3 \times 10^7\,$GeV (astrophobic scenario) and $f_a\sim 2\times 10^9\,$GeV, an axion with a sizeable coupling to electrons ($C_e = 1$).}
    \label{fig:interplayUVIR2}
\end{figure}

The connections established in \eqs{eq:rel1}{eq:rel4} fully prove their relevance once an evidence of deviation from the SM will emerge. Assuming, for example, an axion discovery in a given flavour violating meson decay, it will be then straightforward to relate it to a possible deviation from SM predictions in meson mixing observables, induced by the heavy radial modes of the PQ-2HDM.\footnote{Note that meson mixing receives also an IR contribution from the tree-level exchange of the axion field. However, this contribution is subleading compared to the UV-induced one, as long as $f_a \gg m_H$.}
Fixing the branching ratio of the flavour violating axion observable  
in a region between the current experimental bound and the future expected sensitivity (the latter is also taken from Ref.~\cite{MartinCamalich:2020dfe}),  
we can translate the current bound imposed by meson mixing into a relation of the type $f_a < \#_{f_a} s_{2\beta} m_H$, with $\#_{f_a}$ a number which can be read directly from \eqs{eq:rel1}{eq:rel4} (or equivalently $m_a > \#_{m_a} s_{2\beta} m_H$, using the QCD axion mass relation, $m_a \simeq 5.7 \times (10^6 \ \text{GeV} / f_a)$ eV). This region corresponds the white upper-right half-plane in the plots of \figs{fig:interplayUVIR1}{fig:interplayUVIR2}, where we also superimpose LHC constraints and astrophysical bounds on the axion decay constant. 
In the latter case we assume two scenarios corresponding to a suppressed axion coupling to nucleons and electrons (astrophobic scenario \cite{DiLuzio:2017ogq}) 
and $C_e = 1$ (corresponding to a sizeable value of the axion coupling to electrons). 

If the axion is discovered in the near future (i.e.~in the region between current bounds and future sensitivities) in $B$- or $D$-meson transitions, it is evident from \figs{fig:interplayUVIR1}{fig:interplayUVIR2} that the interplay of meson mixing and astrophysical bounds disfavour the possibility of probing the radial modes of the PQ-2HDM at the LHC (unless the axion is to some extent astrophobic). On the other hand, a discovery of the axion in $K \to \pi a$ leaves open plenty of parameter space for a light PQ-2HDM in the LHC range, since astrophysical constraints are not relevant for large $f_a$. 

A similar discussion applies if a SM deviation will be detected in meson mixing. In such a case \figs{fig:interplayUVIR1}{fig:interplayUVIR2} would be filled in the upper-right plane, and one could draw well-definite conclusions regarding the possibility to observe an axion within the reach of current or future meson decay experiments. This scenario is however more difficult to be realized in practice, since it relies on a precise SM computation for the meson mixing parameters.

\subsection{IR/UV lepton flavour connection}\label{sec:LFV_connection}

In the charged-lepton sector it is possible to exploit \eq{eq:epseC} in order to 
connect LFV axion processes like $l_i\to l_j\, a$ with other LFV observables such as $l_i\to 3l_j$, $l_i\to l_j l_k l_k$ and $l_i\to l_j \gamma$. 
However, as shown in \app{app:LFVobs}, the relation between this two classes of LFV observables is not straightforward as in the quark case considered in \sect{sec:IRUV_meson}, because of the appearance of several couplings which break the 1-to-1 flavour correspondence. Hence, a global analysis  would be necessary in this case.

\begin{figure}[ht]
\centering
\includegraphics[width=0.8\textwidth]{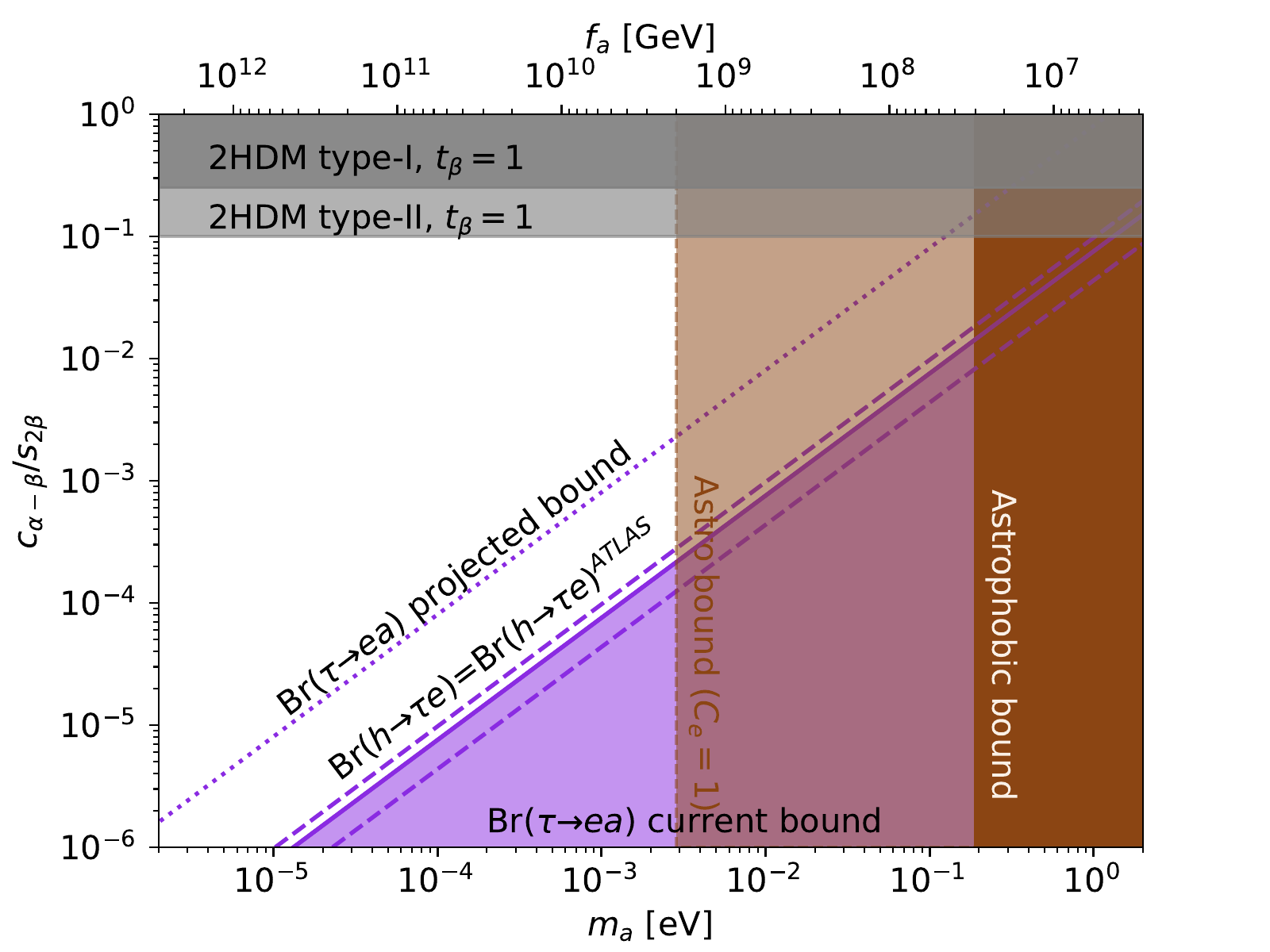} \\ \vspace{0.5cm}
\includegraphics[width=0.8\textwidth]{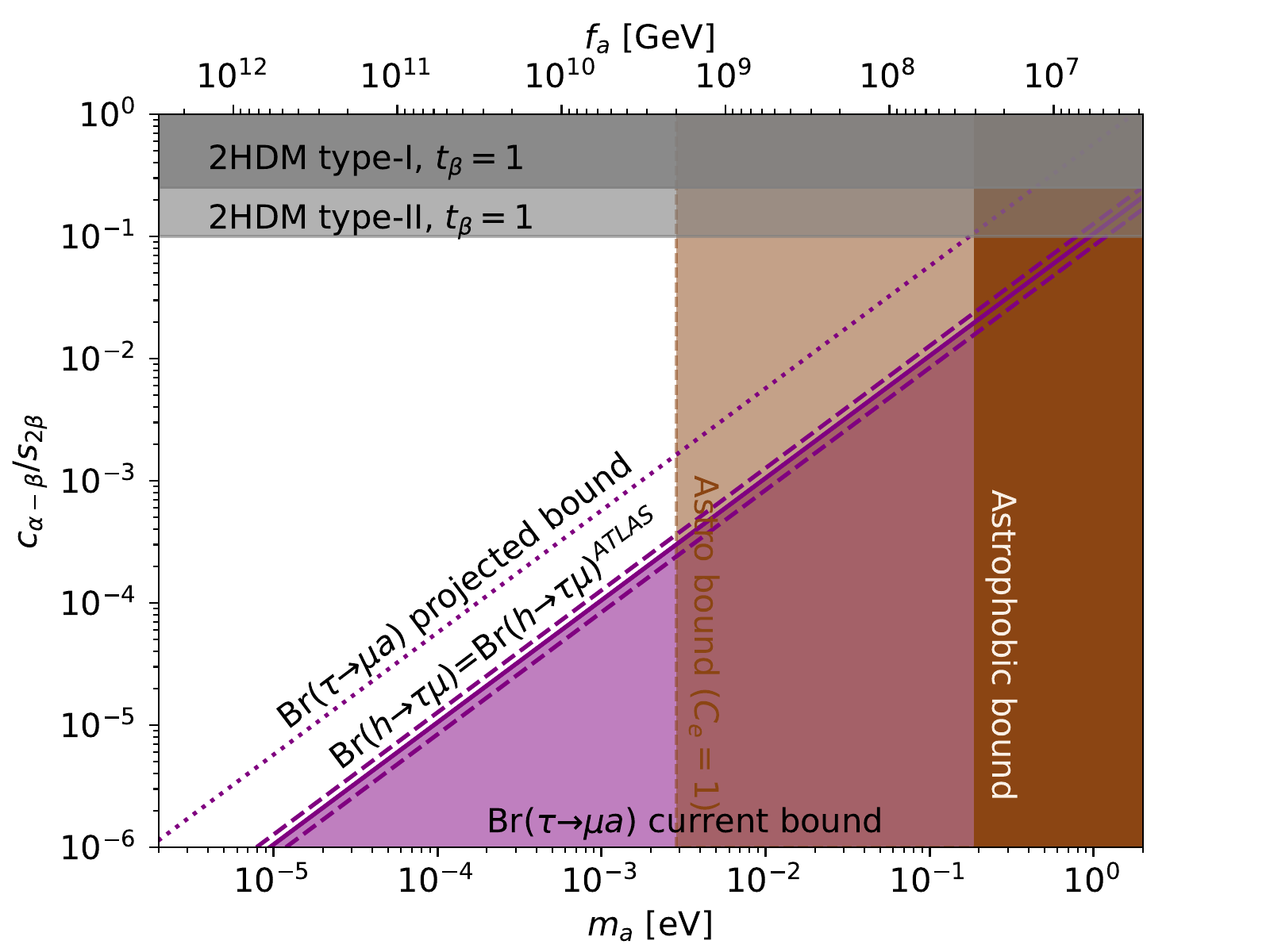}
\caption{IR/UV lepton flavour connection in the $(m_a, c_{\alpha-\beta} / s_{2\beta})$ plane, assuming the hint on 
LFV Higgs decays $h \to \tau \mu/e$ and imposing the bound (colored purple region being excluded) from $\tau\to e a$ 
(top panel) and $\tau\to \mu a$ (bottom panel). Solid (dotted) lines correspond to current bounds (future sensitivities) 
from $\tau$ decays, while dashed lines refer to the 1$\sigma$ region for the $h \to \tau \mu/a$ hint. 
For input data see \Table{tab:LFV_numbers}. LHC bounds on $c_{\alpha-\beta}$ (for $t_\beta = 1$) are displayed in grey
as a reference in the case of the type-I and -II 2HDM, while the brown regions correspond to astrophysical bounds on 
the axion decay constant, between $f_a \sim 3 \times 10^7\,$GeV (astrophobic scenario) and $f_a\sim 2\times 10^9\,$GeV, 
for an axion with a sizeable coupling to electrons ($C_e = 1$).}
\label{fig:LFVcon}
\end{figure}

Another possibility is to consider LFV Higgs decays $h\to l_i l_j$, which actually turns out to be in direct connection with LFV axion processes, thus providing a useful simplification. Moreover, there is a recent hint from ATLAS \cite{ATLAS:2023mvd} about a possible deviation from the SM in the Higgs branching ratios $\text{BR}(h \to \tau e )=0.09\pm 0.06\,\%$ and  $\text{Br}(h \to \tau \mu)=0.11^{+0.05}_{-0.04}\,\%$. These provide indeed a useful benchmark in order to highlight the IR/UV lepton flavour connection.  Note, also, that deviations at the 0.1$\%$ level in $\text{BR}(h \to \tau e/\mu)$ within a non-universal 2HDM are still allowed by other LFV observables (see e.g.~\cite{Harnik:2012pb}). 
Employing \eq{eq:BRhiggsGeneral} and the IR/UV connection formulae in \sect{sec:IRUVconnection}, we obtain the following expression for the LFV Higgs decays in terms of axion couplings
\beq
	\text{BR}(h\to l_i l_j)\simeq \frac{m_h}{16\pi\Gamma_h}\( \frac{c_{\alpha-\beta}}{s_\beta c_\beta}\)^2\frac{m_{l_i}^2}{v^2}|(C^{L,R}_e)_{ij}|^2 \, , 
\eeq
which applies to models with 2+1 PQ charges (the extension to general non-universal DFSZ models, with both LH and RH axion flavour violation, is provided in \eq{eq:generalhLFV}). Given a LFV process involving the axion field
\cite{Calibbi:2020jvd}
\beq
    \text{BR}(l_i\to l_j \, a)\simeq \frac{m_{l_i}^3}{16\pi\Gamma_{l_i}}\frac{|(C^{L,R}_e)_{ij}|^2}{2f_a^2} \, , 
\eeq
we can provide a direct relation for the two branching ratios, that reads 
\beq
\label{eq:corrLFV}
    \text{BR}(h\to l_i l_j)\simeq \text{BR}(l_i\to a \, l_j)\frac{2m_h}{m_{l_i}}\frac{\Gamma_{l_i}}{\Gamma_h}\frac{f_a^2}{v^2}\left(\frac{c_{\alpha-\beta}}{c_\beta s_\beta}\right)^2 \, . 
\eeq
Employing the data in \Table{tab:LFV_numbers}, we display in \fig{fig:LFVcon} 
the correlation provided by \eq{eq:corrLFV} in the $(m_a , c_{\alpha-\beta}/s_{2\beta})$ plane. Assuming as a benchmark the hint from ATLAS on $h \to \tau \mu/e$ 
decays, we impose the bounds and future sensitivities from the axion processes 
$\tau\to e a$ (top panel) and $\tau\to \mu a$ (bottom panel), as well as reference astrophysical limits on the axion mass and LHC constraints on $c_{\alpha-\beta}$ 
for typical model parameters. The combination of all bounds corners a region of the axion mass parameter space that will be targeted by future axion searches 
(such as those with helioscopes) and that is characterized by a remarkable complementarity between different types of observables, ranging from flavour physics and LHC direct searches to axion astrophysics. 
\begin{table}[t!]
    \centering
    \begin{tabular}{c|c | c| c }
            $i,\,j$ &   BR($h\to l_i l_j$) & Current BR($l_i\to a \, l_j$) & Expected BR($l_i\to a l_j$) \\\hline
        $\tau,\, e$  & $0.09\pm 0.06\,\%$ \cite{ATLAS:2023mvd} & $9.4\times 10^{-4}$ \cite{Belle-II:2022heu}& $8.3\times 10^{-6}$ \cite{Calibbi:2020jvd}\\
        $\tau,\, \mu$ & $0.11^{+0.05}_{-0.04}\,\%$  \cite{ATLAS:2023mvd}&$5.9\times 10^{-4}$ \cite{Belle-II:2022heu} & $2.0\times 10^{-5}$ \cite{Calibbi:2020jvd}\\
    \end{tabular}
    \caption{Input values used in the LFV analysis in \fig{fig:LFVcon}. Lepton decays correspond to $95\%$ CL bounds, while the Higgs branching ratios are 
    measurements
    at $1\sigma$.}
    \label{tab:LFV_numbers}
\end{table}

\clearpage
\section{Conclusions}
\label{sec:concl}

In this paper we have studied the flavour phenomenology of non-universal DFSZ axion models, focusing on the interplay 
between IR and UV sources of flavour violation. These consist respectively in the flavour violating axion couplings to 
fermions, whose chiral counterparts are denoted as $C_f^{L,R}$ (see \eq{eq:CRLf}) and the flavour violating couplings of 
the heavy radial modes of the PQ-2HDM, which are expressed in terms of the $\epsilon^f$ parameters (defined in 
\eq{eq:EpsilonCouplings}). The key relations, providing a direct link between the $\epsilon^f$ and $C_f^{L,R}$ coefficients, 
are given by \eqs{eq:epsuC}{eq:epseC}. These remarkably simple equations, that lock the flavour violating pattern of the axion 
field to that of the heavy radial modes, hold in general for any (non-universal) DFSZ models, with a renormalizable 2HDM 
embedded. In fact, in order to derive \eqs{eq:epsuC}{eq:epseC} we have exploited the $\U(1)_{\rm PQ}$ invariance of the 
PQ-2HDM Yukawa sector made explicit in \eqs{eq:chargesu}{eq:chargese}.   

With the help of \eqs{eq:epsuC}{eq:epseC} 
we have hence investigated the flavour phenomenology both in the quark and charged-lepton sectors, also in connection 
with LHC limits on the PQ-2HDM and astrophysical constraints on axion couplings. In the quark sector we focused on 
meson mixing observables, which yield the strongest constraints on the flavour violating couplings of the non-universal 
PQ-2HDM radial modes. We have also derived a set of useful formulae (see \eqs{eq:BoundDecouplingKaon2}{eq:BoundDecouplingD2}), 
from which present bounds on meson mixing 
mass differences can be translated to tree-level flavour violating couplings of a general 2HDM in the alignment limit, regardless of the presence 
of the PQ symmetry. 
  
In order to further exemplify the IR/UV flavour connection we have considered two explicit examples of 2+1 models, labelled 
M1 and M4, which are prototypes of models with quark flavour violation in axion couplings, respectively in the LH and in the RH 
down-quark sector. Assuming an axion discovery in flavour violating meson decays, and imposing in turn the bounds from 
meson mixing experiments we have obatined the results shown in \figs{fig:interplayUVIR1}{fig:interplayUVIR2}. The main 
message that can be learnt with such exercise is that if an axion is discovered in the near future in $B$- or $D$-meson 
transitions, then the interplay of meson mixing and astrophysical bounds disfavours the possibility of probing the radial 
modes of the PQ-2HDM at the LHC. On the other hand, an axion discovery in $K \to \pi a$ leaves open plenty of parameter 
space for a light PQ-2HDM in the LHC range. 

A similar IR/UV connection can be envisaged as well for the charged-lepton sector. However, LFV observables, such as 
for example $l_i \to l_j \gamma$, turn out to depend on several axion couplings, making a direct connection with LFV 
axion processes not straightforward. To this end, we have considered instead SM-like Higgs decays $h \to l_i l_j$, which actually provide a 1-to-1 flavour correspondence with LFV axion processes 
such as $l_i \to l_j a$. Given the fact that ATLAS \cite{ATLAS:2023mvd} currently hints to a mild deviation from the 
SM in the Higgs decays $h \to \tau e/\mu$, we have taken this as a benchmark scenario and imposed the bounds from axion 
LFV processes, exploiting again the relation between IR and UV sources of flavour violation. The results of this analysis 
are displayed in \fig{fig:LFVcon}, which shows again a strong complementarity between flavour observables, LHC direct 
searches and axion astrophysics.

\section*{Acknowledgements}

We thank Mario Fern\'andez Navarro, Bel\'en Gavela, Luca Merlo and Robert Ziegler for useful discussions and comments. 
This work received funding from the European Union's Horizon 2020 research and innovation programme under the Marie 
Sk\l{}odowska-Curie grant agreement N$^{\circ}$ 860881-HIDDeN, grant agreement N$^{\circ}$ 101086085–ASYMMETRY 
and by the INFN Iniziative Specifica APINE. The work of LDL is also supported by the project ``CPV-Axion'' under 
the Supporting TAlent in ReSearch@University of Padova (STARS@UNIPD).

\appendix

\section{PQ-2HDM scalar potential}\label{app:2HDMpotential}
\label{sec:scalpot}

In this Appendix we provide some details on the scalar potential, 
with a special focus on the PQ-2HDM spectrum in the PQ broken phase (see also \cite{Bertolini:2014aia,Espriu:2015mfa}). 
The DFSZ scalar potential, in the presence of the cubic invariant $H_2^\dag H_1 \phi$, 
can be written as 
\begin{align}
\label{eq:VDFSZ}
    V\left( H_1, H_2, \phi \right) &= \mu_1^2 |H_1|^2 +\mu_2^2 |H_2|^2  +\frac{\lambda_1}{2}|H_1|^4 +\frac{\lambda_2}{2}|H_2|^4 
    +\lambda_3|H_1|^2|H_2|^2 \nonumber \\ 
    &+\lambda_4( H_1^\dagger H_2)( H_2^\dagger H_1) +\frac{\lambda^\phi_1 }{2}|\phi|^2 |H_1|^2 +\frac{\lambda^\phi_2 }{2}|\phi|^2 |H_2|^2 
    +\lambda^\phi_3 \(|\phi|^2 - \frac{v^2_\phi}{2}\)^2
    \nonumber \\ 
    &- \left(\mu_\phi H^\dagger_2 H_1 \phi + \text{h.c.}\right) \, ,
\end{align}  
where the
Higgs doublets are decomposed as 
\beq
\label{eq:defH12}
H_1 = 
\begin{pmatrix}
\frac{1}{\sqrt{2}} (v_1 + h^0_{1} + i \eta_1) \\
h^-_1
\end{pmatrix} \, , \quad  
H_2 = 
\begin{pmatrix}
\frac{1}{\sqrt{2}} (v_2 + h^0_{2} + i \eta_2) \\
h^-_2
\end{pmatrix} \, . \quad 
\eeq
In the PQ-broken phase, with $\langle \phi\rangle=v_{\phi}/\sqrt{2}$, we can integrate out 
the radial mode of the $\phi$ field and obtain the leading-order PQ-2HDM potential 
\begin{align}
    V 
    \left( H_1, H_2 \right) &= m_{11}^2 |H_1|^2 +m_{22}^2 |H_2|^2 - \left(m_{12} H^\dagger_1 H_2 + \text{h.c.}\right)  \nonumber \\ 
    &+\frac{\lambda_1}{2} |H_1|^4 +\frac{\lambda_2}{2} |H_2|^4 +\lambda_3 |H_1|^2 |H_2|^2 
    +\lambda_4( H_1^\dagger H_2)( H_2^\dagger H_1) \, , 
\end{align}
where 
\begin{equation}
    m_{11}^2=\mu^2_1+\frac{\lambda_1^\phi}{4}v^2_{\phi} \, , \qquad 
    m_{22}^2=\mu^2_2+\frac{\lambda_2^\phi}{4}v^2_{\phi}\, , \qquad 
    m_{12}^2=\mu^*_{\phi} \frac{v_{\phi}}{\sqrt{2}} \, .
\end{equation}
To implement the $v_\phi \gg v$ hierarchy, we impose the ``ultra-weak'' scaling 
$\lambda_{1,2}^\phi\sim v^2/v^2_{\phi}$ and $\mu_\phi\sim v^2/v_{\phi}$, which also 
guarantees that the 
hierarchy between the 
PQ and the electroweak breaking scales 
remains radiatively stable thanks to the emergence of an extended 
Poincar\'e symmetry of the action \cite{Volkas:1988cm,Foot:2013hna}.  
This is reflected by the fact that $\lambda_{1,2}^\phi \to 0$ and $\mu_\phi \to 0$  
correspond to fixed points of the renormalization group evolution (see e.g.~App.~B in \cite{Bertolini:2015boa}).

To match with the more standard 2HDM notation (see e.g.~\cite{Gunion:2002zf})
we consider the rotated doublets 
\beq 
\label{eq:rotation_beta}
\begin{pmatrix}
\Phi_1 \\ \Phi_2 \\
\end{pmatrix}
= 
\begin{pmatrix}
c_\beta & s_\beta \\ 
- s_\beta & c_\beta 
\end{pmatrix} 
\begin{pmatrix}
\tilde H_1 \\ \tilde H_2 \\
\end{pmatrix} \, ,
\eeq
so that $\Phi_1$ picks up the whole electroweak VEV 
\beq
\label{eq:rotatedfields1}
\Phi_1 = 
\begin{pmatrix}
G^+ \\ 
- \frac{1}{\sqrt{2}} (v + c_\beta h^0_{1} + s_\beta h^0_{2} - i G^0) 
\end{pmatrix} \, , \quad  
\Phi_2 = 
\begin{pmatrix}
H^+ \\ 
- \frac{1}{\sqrt{2}} (- s_\beta h^0_{1} + c_\beta h^0_{2} - i A) \end{pmatrix} \, .  
\eeq
Here we have defined the would-be Goldstone bosons 
\beq 
G^+ = c_\beta h^+_1 + s_\beta h^+_2 \, , \quad 
G^0 = c_\beta \eta_1 + s_\beta \eta_2 \, ,
\eeq 
and the scalar radial modes 
\beq 
H^+ = -s_\beta h^+_1 + c_\beta h^+_2 \, , \quad 
A^0 = -s_\beta \eta_1 + c_\beta \eta_2 \, .
\eeq  
In the rotated basis the scalar potential reads 
\begin{align}
        V\left( \Phi_1, \, \Phi_2 \right) &= M_{11}^2 \,\Phi_1^\dagger \Phi_1 +M_{22}^2\, \Phi_2^\dagger \Phi_2  - \left(M_{12} \Phi_1^\dagger \Phi_2 + \text{h.c.}\right)  \nonumber \\ &+\frac{\Lambda_1}{2}( \Phi_1^\dagger \Phi_1)^2 +\frac{\Lambda_2}{2}( \Phi_2^\dagger \Phi_2)^2 +\Lambda_3( \Phi_1^\dagger \Phi_1)( \Phi_2^\dagger \Phi_2) +\Lambda_4( \Phi_1^\dagger \Phi_2)( \Phi_2^\dagger \Phi_1) \nonumber \\ 
        &+\left\{\frac{1}{2}\Lambda_5 (\Phi_1^\dagger \Phi_2)^2 + \left[\Lambda_6(\Phi_1^\dagger \Phi_1)+\Lambda_7(\Phi_2^\dagger\Phi_2)\right]\Phi_1^\dagger\Phi_2+\text{h.c.}\right\} \, , 
\end{align}
with the new parameters given by
\begin{align}
    M_{11}^2&=m_{11}^2 c_\beta^2+m_{22}^2 s_\beta^2 -\Re{m_{12}^2}\, s_{2\beta} \, , \\
    M_{22}^2&=m_{11}^2 s_\beta^2+m_{22}^2 c_\beta^2 -\Re{m_{12}^2} \,s_{2\beta} \, , \\
    M_{12}^2&=\frac{1}{2}\left(m_{11}^2 -m_{22}^2 \right)s_{2\beta} +\Re{m_{12}^2}\,c_{2\beta} +i\,\Im{m_{12}^2} \, , \\
    \Lambda_1&=\lambda_1 c^4_\beta+\lambda_2 s^4_\beta+ \frac{1}{2}(\lambda_3+\lambda_4)s^2_{2\beta} \, , \\
     \Lambda_2&=\lambda_1 s^4_\beta+\lambda_2 c^4_\beta+ \frac{1}{2}(\lambda_3+\lambda_4)s^2_{2\beta} \, , \\
     \Lambda_3&=\frac{1}{4}s^2_{2\beta}\left[\lambda_1+\lambda_2 - 2(\lambda_3+\lambda_4)\right]+\lambda_3 \, , \\
     \Lambda_4&=\frac{1}{4}s^2_{2\beta}\left[\lambda_1+\lambda_2 - 2(\lambda_3+\lambda_4)\right]+\lambda_4 \, , \\
     \Lambda_5&=\frac{1}{4}s^2_{2\beta}\left[\lambda_1+\lambda_2 - 2(\lambda_3+\lambda_4)\right] \, , \\
     \Lambda_6&=-\frac{1}{2}s_{2\beta}\left[\lambda_1 c_\beta^2-\lambda_2 s_\beta^2 -c_{2\beta}(\lambda_3+\lambda_4)\right]  \, , \\
      \Lambda_7&=-\frac{1}{2}s_{2\beta}\left[\lambda_1 s_\beta^2-\lambda_2 c_\beta^2 +c_{2\beta}(\lambda_3+\lambda_4)\right]  \, .
\end{align}
The stationary conditions read
\begin{align}
   \left. \frac{\partial V}{\partial \Phi_1}\right\vert_{\langle\Phi_2\rangle=0,\,\langle\Phi_1\rangle=\frac{v}{\sqrt{2}}}= 0 \, , \quad &\Longrightarrow \quad
   M_{11}^2\frac{v}{\sqrt{2}}+\Lambda_1\frac{v^3}{2\sqrt{2}}
   = 0 
   \, , \\
   \left. \frac{\partial V}{\partial \Phi_2}\right\vert_{\langle\Phi_2\rangle=0,\,\langle\Phi_1\rangle=\frac{v}{\sqrt{2}}}= 0 \, , \quad &\Longrightarrow \quad
   -M_{12}^2\frac{v}{\sqrt{2}}+\Lambda_6\frac{v^3}{2\sqrt{2}}
   = 0
   \, , 
\end{align}
from which we can substitute $M^2_{11}$ and $M^2_{12}$ 
into the scalar potential. 
The squared mass terms of the CP-odd and charged scalars can be read directly from the Lagrangian
\begin{align}
    m_{H^\pm}^2&=M^2_{22}+\frac{1}{2}v^2\Lambda_3 \, , \\
    m_A^2&=m^2_{H^{\pm}}+\frac{1}{2}v^2(\Lambda_4-\Lambda_5)=M^2_{22}+\frac{1}{2}v^2(\Lambda_3+\Lambda_4-\Lambda_5) \, , 
\end{align}
while to obtain those of the
CP-even Higgses we perform the rotation 
\beq 
\label{eq:rotation_alpha}
\begin{pmatrix}
h^0_{2} \\ h^0_{1} \\
\end{pmatrix}
= 
\begin{pmatrix}
c_\alpha & s_\alpha \\ 
- s_\alpha & c_\alpha 
\end{pmatrix} 
\begin{pmatrix}
h \\ H \\
\end{pmatrix} \, .
\eeq
This is equivalent to go into the
basis where only one Higgs gets the whole electroweak VEV  
\begin{equation}
    \label{eq:CPevenDiagMasses}
    \begin{pmatrix}
    m_H^2 && 0 \\
    0 && m_h^2
    \end{pmatrix}=\begin{pmatrix}
    c_{\alpha-\beta} && s_{\alpha-\beta} \\
    -s_{\alpha-\beta} && c_{\alpha-\beta}
    \end{pmatrix}\begin{pmatrix}
    \Lambda_1 v^2 && \Lambda_6 v^2 \\
    \Lambda_6 v^2 && m_A^2+\Lambda_5 v^2
    \end{pmatrix}\begin{pmatrix}
    c_{\alpha-\beta} && -s_{\alpha-\beta} \\
    s_{\alpha-\beta} && c_{\alpha-\beta}
    \end{pmatrix} \, ,
\end{equation}
with eigenvalues
\begin{equation}
\label{eq:CPevenmasses}
    m_{h,\,H}^2=\frac{1}{2}\left(m_A^2+v^2(\Lambda_1+\Lambda_5)\mp\sqrt{\left[m_A^2+(\Lambda_5-\Lambda_1)v^2\right]^2+4v^4\Lambda_6^2}\right) \, ,
\end{equation}
and mixing angle given by \cite{Gunion:2002zf}
\begin{equation}
    \label{eq:sin2alphabeta}
    s_{2(\alpha-\beta)}=\frac{2\Lambda_6 v^2}{m_H^2-m_h^2}, \,\,\,\,\,\,\,\, c_{2(\alpha-\beta)}=\frac{(\Lambda_1-\Lambda_5)v^2-m_A^2}{m_H^2-m_h^2} \, .
\end{equation}
A useful quantity (appearing into Higgs couplings) is 
\begin{equation}
    \label{eq:cos2alphabeta}
    c^2_{\alpha-\beta}=\frac{\Lambda_1 v^2 -m_h^2}{m_H^2-m_h^2} 
    = 
    \frac{\Lambda^2_6 v^4}{(m_H^2-m_h^2) (m_H^2-\Lambda_1 v^2)}
    \, ,  
\end{equation}
while some of the previous equations can be rearranged as 
\begin{align}
    \label{eq:massrelations1}
    \Lambda_1 v^2 & = m_h^2 s_{\alpha-\beta} ^2+m_H^2 c_{\alpha-\beta}^2 \, , \\
      \label{eq:massrelations2}
    \Lambda_6 v^2 & = ( m_h^2 -m_H^2 )\, s_{\alpha-\beta} c_{\alpha-\beta} \, , \\
      \label{eq:massrelations3}
    m_A^2 + \Lambda_5 v^2 & = m_H^2 s_{\alpha-\beta}^2+m_h^2 c_{\alpha-\beta}^2 \, .
\end{align}
Finally, \eq{eq:PQHiggsdoublets} provides the embedding 
of the physical mass eigenstates in the 
original basis of the PQ-Higgs doublets. 

Note that $s_{\alpha-\beta} = 0$ corresponds to $\Lambda_6 = 0$ . 
In this scenario 
the field $H$ is \textit{aligned} to the electroweak VEV, making $H$ to be SM-like with $m^2_H=\Lambda_1 v^2$ \cite{Bernon:2015qea}. In the opposite case where 
$c_{\alpha-\beta}=0$ we have instead 
$m_h^2=\Lambda_1 v^2$, 
corresponding to a SM-like $h$.
This can be obtained by taking $m_H^2\gg v^2$, 
which from Eq.~(\ref{eq:cos2alphabeta}) 
that corresponds to $c_{\alpha-\beta}\sim 1/m_H^2$. 
This scenario is called \textit{alignment with decoupling} \cite{Bernon:2015wef}. From \eq{eq:massrelations3} 
one also obtains 
\begin{equation}
    \label{eq:Lambda5}
    m_A^2+\Lambda_5 v^2 \simeq m_H^2 \, .
\end{equation}
Hence, in the decoupling scenario the scalar masses become degenerate, $m_H^2\simeq m_A^2 \simeq m_{H^\pm}^2$. This is instead not the case for the alignment without decoupling, where the scalar masses remain at the electroweak scale and, in general, are non-degenerate. Such a case requires  $\Lambda_6\to 0$ and it can lead to a non-SM-like Higgs 
either above or below $125\,$GeV, see \cite{Bernon:2015qea,Bernon:2015wef} respectively.

\section{4-fermion operators}
\label{sec:4fermion}

We list here the 4-fermion operators which arise in \eq{eq:def_4fermion}, upon integrating out at tree level the 
radial modes of the PQ-2HDM.  In the quark sector we obtain
\begin{align}
\mathcal{L}^{\rm 4\text{-}quark}_{\rm EFT} &= 
( C_{ijkl}^{\bar u_L u_R \bar u_L u_R} \bar u_{Li} u_{Rj} \bar u_{Lk} u_{Rl} + \text{h.c.} )+( C_{ijkl}^{\bar d_L d_R \bar d_L d_R} \bar d_{Li} d_{Rj} \bar d_{Lk} d_{Rl} + \text{h.c.} ) \nonumber \\
&+( C_{ijkl}^{\bar u_L u_R \bar d_L d_R} \bar u_{Li} u_{Rj} \bar d_{Lk} d_{Rl} + \text{h.c.} )+ ( C_{ijkl}^{\bar u_L u_R \bar d_R d_L} \bar u_{Li} u_{Rj} \bar d_{Rk} d_{Ll} + \text{h.c.} ) \nonumber \\
&+  C_{ijkl}^{\bar u_L u_R \bar u_R u_L} \bar u_{Li} u_{Rj} \bar u_{Rk} u_{Ll} +  C_{ijkl}^{\bar d_L d_R \bar d_R d_L} \bar d_{Li} d_{Rj} \bar d_{Rk} d_{Ll}  \nonumber \\
&+ ( C_{ijkl}^{\bar u_L d_R \bar d_L u_R} \bar u_{Li} d_{Rj} \bar d_{Lk} u_{Rl} + \text{h.c.} ) 
+ C_{ijkl}^{\bar u_L d_R \bar d_R u_L} \bar u_{Li} d_{Rj} \bar d_{Rk} u_{Ll} \nonumber \\
&+ C_{ijkl}^{\bar d_L u_R \bar u_R d_L} \bar d_{Li} u_{Rj} \bar u_{Rk} d_{Ll} \, , 
\label{eq:4quarks}
\end{align} 
where the 4-quark Wilson coefficients are found to be
\begin{align}
C^{\bar u_L u_R \bar u_L u_R}_{ijkl} &= \frac{1}{2 m_H^2} C^{H_u}_{ij} C^{H_u}_{kl} + \frac{1}{2 m_A^2} C^{A_u}_{ij} C^{A_u}_{kl}+ \frac{1}{2 m_h^2} C^{h_u}_{ij} C^{h_u}_{kl} \, , \\ 
C^{\bar u_L u_R \bar d_L d_R}_{ijkl} &= \frac{1}{2 m_H^2} C^{H_u}_{ij} C^{H_d}_{kl} + \frac{1}{2 m_A^2} C^{A_u}_{ij} C^{A_d}_{kl}+ \frac{1}{2 m_h^2} C^{h_u}_{ij} C^{h_d}_{kl} \, , \\
C^{\bar d_L d_R \bar d_L d_R}_{ijkl} &= \frac{1}{2 m_H^2} C^{H_d}_{ij} C^{H_d}_{kl} + \frac{1}{2 m_A^2} C^{A_d}_{ij} C^{A_d}_{kl}+ \frac{1}{2 m_h^2} C^{h_d}_{ij} C^{h_d}_{kl} \, , \\
C^{\bar u_L u_R \bar d_R d_L}_{ijkl} &= \frac{1}{2 m_H^2} C^{H_u}_{ij} (C^{H_d}_{lk} )^*+ \frac{1}{2 m_A^2} C^{A_u}_{ij} (C^{A_d}_{lk})^*+ \frac{1}{2 m_h^2} C^{h_u}_{ij} (C^{h_d}_{lk})^* \, , \\ 
C^{\bar u_L u_R \bar u_R u_L}_{ijkl} &= \frac{1}{ m_H^2} C^{H_u}_{ij} (C^{H_u}_{lk})^* + \frac{1}{ m_A^2} C^{A_u}_{ij} (C^{A_u}_{lk})^*+ \frac{1}{ m_h^2} C^{h_u}_{ij} (C^{h_u}_{lk})^* \, , \\
C^{\bar d_L d_R \bar d_R d_L}_{ijkl} &= \frac{1}{ m_H^2} C^{H_d}_{ij} (C^{H_d}_{lk} )^*+ \frac{1}{ m_A^2} C^{A_d}_{ij} (C^{A_d}_{lk})^*+ \frac{1}{ m_h^2} C^{h_d}_{ij} (C^{h_d}_{lk})^* \, , \\
C^{\bar u_L d_R \bar d_L u_R}_{ijkl} &= \frac{2}{ m_{H^\pm}^2} V_{nj} V^*_{mk} C^{H^+_d}_{in} C^{H^-_u}_{ml} \, ,  \\ 
C^{\bar u_L d_R \bar d_R u_L}_{ijkl} &= \frac{2}{ m_{H^\pm}^2} V_{nj} V_{mk}^* C^{H^+_d}_{in} (C^{H^+_d}_{lm})^*  \, , \\
C^{\bar d_L u_R \bar u_R d_L}_{ijkl} &= \frac{2}{ m_{H^\pm}^2} V^*_{ni} V_{ml} C^{H^-_u}_{nj} (C^{H^-_u}_{mk} )^* \, .
\end{align}
 For the semi-leptonic operators we find 
 \begin{align}
 \label{eq:lag_semi}
\mathcal{L}^{\rm semi\text{-}lept}_{\rm EFT} &= 
( C_{ijkl}^{\bar q_L q_R \bar e_L e_R} \bar q_{Li} q_{Rj} \bar e_{Lk} e_{Rl} + \text{h.c.} )
+ ( C_{ijkl}^{\bar q_L q_R \bar e_R e_L} \bar q_{Li} q_{Rj} \bar e_{Rk} e_{Ll} + \text{h.c.} ) \nonumber \\
&+ ( C_{ijkl}^{\bar \nu_L e_R \bar d_R u_L} \bar \nu_{Li} e_{Rj} \bar d_{Rk} u_{Ll} + \text{h.c.} ) + ( C_{ijkl}^{\bar \nu_L e_R \bar d_R u_L} \bar \nu_{Li} e_{Rj} \bar d_{Rk} u_{Ll} + \text{h.c.} ) \, ,
\end{align} 
 with Wilson coefficients
 \begin{align}
 C^{\bar u_L u_R \bar e_L e_R}_{ijkl} &= \frac{1}{ m_H^2} C^{H_u}_{ij} C^{H_e}_{kl} + \frac{1}{ m_A^2} C^{A_u}_{ij} C^{A_e}_{kl}+ \frac{1}{m_h^2} C^{h_u}_{ij} C^{h_e}_{kl} \, , \\ 
C^{\bar d_L d_R \bar e_L e_R}_{ijkl} &= \frac{1}{ m_H^2} C^{H_d}_{ij} C^{H_e}_{kl} + \frac{1}{ m_A^2} C^{A_d}_{ij} C^{A_e}_{kl}+ \frac{1}{m_h^2} C^{h_d}_{ij} C^{h_e}_{kl} \, , \\
C^{\bar u_L u_R \bar e_R e_L}_{ijkl} &= \frac{1}{m_H^2} C^{H_u}_{ij} (C^{H_e}_{lk} )^* + \frac{1}{ m_A^2} C^{A_u}_{ij} (C^{A_e}_{lk} )^*+ \frac{1}{ m_h^2} C^{h_u}_{ij} (C^{h_e}_{lk} )^* \, , \\
C^{\bar d_L d_R \bar e_R e_L}_{ijkl} &= \frac{1}{ m_H^2} C^{H_d}_{ij} (C^{H_e}_{lk} )^*+ \frac{1}{ m_A^2} C^{A_d}_{ij} (C^{A_e}_{lk})^*+ \frac{1}{ m_h^2} C^{h_d}_{ij} (C^{h_e}_{lk})^* \, , \\ 
C^{\bar \nu_L e_R \bar d_L u_R}_{ijkl} &= \frac{2}{ m_{H^\pm}^2} U^*_{ni} V_{mk}^* C^{H^+_e}_{nj} C^{H^-_u}_{ml} ,  \\ 
C^{\bar \nu_L e_R \bar d_R u_L}_{ijkl} &= \frac{2}{ m_{H^\pm}^2} U^*_{ni} V_{mk}^* C^{H^-_e}_{nj} (C^{H^+_d}_{lm})^* \, .
 \end{align}
% \LDL{[semi-leptonic operators can be bounded from $\mu \to e \gamma$ after closing the quark loop]}
 Finally, the purely leptonic effective operators are
 \begin{align}
 \mathcal{L}^{\rm 4\text{-}lept}_{\rm EFT} &= 
( C_{ijkl}^{\bar e_L e_R \bar e_L e_R} \bar e_{Li} e_{Rj} \bar e_{Lk} e_{Rl} + \text{h.c.} )
+ C_{ijkl}^{\bar e_L e_R \bar e_R e_L} \bar e_{Li} e_{Rj} \bar e_{Rk} e_{Ll}  \nonumber \\
&+  C_{ijkl}^{\bar \nu_L e_R \bar e_R \nu_L} \bar \nu_{Li} e_{Rj} \bar e_{Rk} \nu_{Ll} \, , 
 \end{align}
 with Wilson coefficients
  \begin{align}
 C^{\bar e_L e_R \bar e_L e_R}_{ijkl} &= \frac{1}{2 m_H^2} C^{H_e}_{ij} C^{H_e}_{kl} + \frac{1}{ 2m_A^2} C^{A_e}_{ij} C^{A_e}_{kl}+ \frac{1}{2m_h^2} C^{h_e}_{ij} C^{h_e}_{kl} \, , \\ 
C^{\bar e_L e_R \bar e_R e_L}_{ijkl} &= \frac{1}{m_H^2} C^{H_e}_{ij} (C^{H_e}_{lk} )^* + \frac{1}{ m_A^2} C^{A_e}_{ij} (C^{A_e}_{lk} )^*+ \frac{1}{ m_h^2} C^{h_e}_{ij} (C^{h_e}_{lk} )^* \, , \\
C^{\bar \nu_L e_R \bar e_R \nu_L}_{ijkl} &= \frac{2}{ m_{H^\pm}^2} U^*_{ni} U_{ml} C^{H^+_e}_{nj} (C^{H^+_e}_{mk})^* \, .
 \end{align}

\section{Meson mixing bounds}
\label{sec:mesonmixing}

The neutral current 4-quark operators in \eq{eq:4quarks} contribute to $\Delta F=2$ transitions, which are constrained by meson mixing observables. 
The effective $\Delta F=2$ Hamiltonian is given by 
\begin{equation}
\mathcal{H}^{\Delta F =2}_{\text{eff}}=\sum_{i}^5 C_i Q_i+ \sum_i^5 \tilde{C}_i \tilde{Q}_i \, ,
    \label{eq:F2Hamilt}
\end{equation}
with 
\begin{align}
    Q_1 &= \bar q^\alpha_L \gamma^\mu q'^\alpha_L  \bar q^\beta_L \gamma_\mu q'^\beta_L \, , \\
    Q_2 &= \bar q^\alpha_R q'^\alpha_L  \bar q^\beta_R q'^\beta_L \, , \\
    Q_3 &=\bar q^\alpha_R q'^\beta_L  \bar q^\beta_R q'^\alpha_L \, , \\
    Q_4 &= \bar q^\alpha_R q'^\alpha_L  \bar q^\beta_L q'^\beta_R \,, \\
    Q_5 &=\bar q^\alpha_R q'^\beta_L  \bar q^\beta_L q'^\alpha_R \, .
\end{align}
Here, $\alpha$ and $\beta$ are color indices, while $q$ and $q'$ refer to two different quark flavours in the mesonic system. 
In the 2HDM case, that is relevant for our analysis, only the operators $Q_2$ and $Q_4$ are generated at tree level. 
However, it is well-known that QCD running effects induce other operators, 
and give a sizable contribution to those already present. 
Running effects will be taken 
into account by using the analytic formulas provided in Refs.~\cite{Ciuchini:1998ix,Becirevic:2001jj,UTfit:2007eik}, 
respectively for $K$, $B$ and $D$ meson mixing. 
The master formula encoding the running of the Wilson coefficients is given by
\begin{equation}
    C_r(\mu)=\sum_i \sum_s \left(b_i^{(r,s)}+\eta\, c^{(r,s)}_i\right)\eta^{a_i} \,C_i(M_H) \, ,
\end{equation}
where, $a, \, b,\, c$ are so-called ``magic numbers'', which depend on the process at hand 
and are taken from the above-mentioned references. The running of the strong coupling is 
taken care by the factor $\eta=\alpha_s(m_H)/\alpha_s(M_Z)$, where we take as an imput $\alpha_s (M_Z)$ 
and compute the QCD running at 2-loops up to the scale of the heavy NP, $m_H$,
by means of the code RunDec \cite{Herren:2017osy}.
The low-energy scale $\mu$ is set to $2$, $4.6$ and $2.8$ GeV respectively for the $K$, $B$ and $D$ meson systems, 
in order to match with the scale of the corresponding hadronic matrix elements.  
The numerical inputs for the meson mixing analysis are given in \Table{tab:meson_mixing}, where we defined
the QCD running coefficients $\eta_i = C_i(\mu) / C_i(m_H)$ and $\eta_{ij} = C_i(\mu) / C_j(m_H)$, and we have taken as a reference value $m_H=1$ TeV. 
\begin{table}[t]
    \centering
    \begin{tabular}{c|c|c|c|c}
         & $K^0-\bar K^0$ & $B^0-\bar B^0$ & $B^0_s-\bar B^0_s$& $D-\bar D$  \\ 
         \hline
         $ \Delta m\,$[GeV]$\times 10^{-15}$ & $3.484\pm 0.006 $ &$ 333.4\pm 1.3 $  &$11693\pm 8 $ & $6.7\pm 0.8 $ \\
         $M_M\,[\text{GeV}]$  & 0.497 & 5.279 & 5.367 & 1.864 \\
         $f_M\,[\text{GeV}]$ & 0.156 & 0.192 & 0.228 & 0.209 \\
         $B_2$ & 0.50 & 0.77 & 0.82  & 0.66\\
         $B_4$  & 0.92 & 1.08 & 1.03 & 0.84\\
         $\eta_{2}$ & 2.50 & 1.99 & 1.99 & 2.30 \\ 
         $\eta_{32}$ & $-1.70\times 10^{-3}$ & $-0.024$ & $-0.024$ & $-0.013$ \\ 
         $\eta_{4}$& 4.78 & 3.10 & 3.10 & 3.97 \\ 
         $\eta_{45}$ & 0.18 & 0.086 & 0.086 & 0.14
    \end{tabular}
    \caption{Numerical inputs for the analysis of $\Delta F = 2$ observables, 
    including  
    experimental values for meson-antimeson mass differences 
    and meson masses \cite{ParticleDataGroup:2020ssz}, 
    meson decay constants \cite{FlavourLatticeAveragingGroupFLAG:2021npn},  
    bag parameters for \Kmix \cite{FlavourLatticeAveragingGroupFLAG:2021npn}, 
    \Bdmix and \Bsmix \cite{Dowdall:2019bea}, \Dmix \cite{Carrasco:2014uya}, 
    as well as  
    QCD running coefficients (see text). 
    }
    \label{tab:meson_mixing}
\end{table}
Taking into account the renormalization group evolution of the Wilson coefficient for $\Delta F = 2$ 
transition, the meson mixing amplitudes read \cite{Ciuchini:1998ix}
\begin{align}
    M_{ij}=\frac{\bra{X} \mathcal{H}^{\Delta F=2}_{\text{eff}}\ket{\bar X}}{2 M_X} &= M_X f_X^2\left[-\frac{5}{24} \frac{M_X^2}{\left(m^q_i+m^q_j\right)^2} \eta_2 B_2(\mu)\left((C_2)_{ij}+(C_2^*)_{ji}\right) \right. \nonumber \\ 
    & \left. +\left(\frac{1}{24}+\frac{1}{4}\frac{M_X^2}{\left(m^q_i+m^q_j\right)^2}\right)\eta_4 B_4(\mu)(C_4)_{ij}\right] \, , 
    \label{eq:amplitude_general}
\end{align}
where $B_i$ are the so-called bag parameters which are calculated in lattice QCD (see \Table{tab:meson_mixing} for the employed numerical values).
Including also the contribution of NP, 
the meson-antimeson mass difference is given by 
(neglecting CP-violating phases)
\begin{equation}
    \label{eq:ratio_exp_SM}
    \Delta m = 2| M_{12}^{\text{SM}}+M_{12}^{\text{NP}} | \, .
\end{equation}
Taking into account that the new physics contribution can be either positive or negative, and that the SM uncertainty is much larger than the experimental one 
(which is henceforth neglected), 
we conservatively consider the 
$2\sigma$ bounds:
\begin{align}
    \label{eq:PositiveBound}
    \Delta m^{exp} = \Delta m^{\text{SM}}-2\sigma_{\Delta m}^{\text{SM}}+2 | M_{12}^{\text{NP}}| \,, \quad \text{for} \quad M_{12}>0 \, , \\
    \label{eq:NegativeBound}
    \Delta m^{exp} = \Delta m^{\text{SM}}+2\sigma_{\Delta m}^{\text{SM}}-2 | M_{12}^{\text{NP}}| \,,  \quad \text{for} \quad M_{12}<0 \, . 
\end{align}
Regarding theoretical predictions,
for the \Bdmix and \Bsmix mixing we will employ the values $\Delta m_d=3.51^{+0.15}_{-0.24}\times 10^{-13}\,$GeV 
and $\Delta m_s=1.211^{+0.046}_{-0.079}\times 10^{-11}\,$GeV
from Ref.~\cite{DiLuzio:2019jyq}, which are obtained by averaging lattice QCD and sum rules results. 
For \Kmix mixing we take $\Delta m_K =(3.3\pm1.2)\times 10^{-15}\,\text{GeV}$, 
that is the
FLAG21 value \cite{FlavourLatticeAveragingGroupFLAG:2021npn} (see also \cite{Brod:2011ty}).
For \Dmix mixing a reliable SM prediction is still lacking, and hence 
we will just impose that new physics does not overshoot the experimental value at $2\sigma$.
Taking all of the above into account and applying \eqs{eq:PositiveBound}{eq:NegativeBound} we obtain
\begin{align}
    \left\vert-0.16\,\left((C^d_2)_{12}+(C^d_2)^*_{21}\right)+0.69\,(C^d_4)_{12}\right\vert = \left[2.0, \, 2.5\right] \cdot 10^{-15} \,\text{GeV} \label{eq:boundK}\, , \\
    \left\vert-0.20\,\left((C^d_2)_{13}+(C^d_2)^*_{31}\right)+0.57\,(C^d_4)_{13}\right\vert  = \left[4.6, \,3.0\right] \cdot 10^{-14}\,\text{GeV} \label{eq:boundBd}\, , \\
    \left\vert-0.28\,\left((C^d_2)_{23}+(C^d_2)^*_{32}\right)+0.82\,(C^d_4)_{23}\right\vert = \left[ 1.3, \, 1.2\right]\cdot 10^{-12}\,\text{GeV} \label{eq:boundBs}\, ,
\end{align}
where the number to the left is for the case in which the 
new physics contribution is negative, 
while the number on the right is for the case in which it is positive. 
For $D$ mixing instead the bound reads
\begin{equation}
        \left \vert -0.16\,\left((C^u_2)_{12}+(C^u_2)^*_{21}\right)+0.53\,(C^u_4)_{12} \right \vert = 8.3\cdot 10^{-15}\,\text{GeV} \label{eq:boundD}\, . 
\end{equation}
In this Appendix we have fixed as a reference the running of the $d$($u$)-type sector to $1(10)$ TeV, 
while in the numerical analysis in \sect{sec:pheno} we have properly taken into account running effects as a function of $m_H$.
Note that employing  model-independent constraints on individual Wilson coefficients would lead to an incorrect bound for the 2HDM case, since a specific combination of Wilson coefficients is actually constrained. 

The operators associated to the Wilson coefficients $C_2^f$ and $C_4^f$ are obtained from \eq{eq:4quarks}, 
with the conditions $i=k$, $j=l$ and $i\neq j$. 
Using this information, one can rewrite the Wilson coefficients in terms of $\epsilon_{ij}^f$ 
as follows 
\begin{align}
\label{eq:C2d}
	C_2^d&=\frac{(\epsilon_{ij}^{d})^2}{4c_\beta^2}\(\frac{m_A^2-m_H^2}{m_H^2 m_A^2}+c_{\alpha-\beta}^2\frac{m_H^2-m_h^2}{m_H^2 m_h^2} \) \, , \\
\label{eq:C4d}
	C_4^d&=\frac{\epsilon_{ij}^{d\,}\epsilon_{ji}^{f\, *}}{2c_\beta^2}\(\frac{m_A^2+m_H^2}{m_H^2 m_A^2}+c_{\alpha-\beta}^2\frac{m_H^2-m_h^2}{m_H^2 m_h^2} \) \, ,
\end{align}
and similarly for the up-quark sector, which is obtained by replacing $c_\beta \to s_\beta$ in the above formulae. Since $c_{\alpha-\beta}$ is 
bounded by Higgs observables,  
the second term in \eqs{eq:C2d}{eq:C4d} is somewhat suppressed.
In fact, this becomes obvious in the case of alignment, $c_{\alpha-\beta}\simeq0$, with and without decoupling. 
In the alignment scenarios the Wilson coefficients can be further simplified as 
\begin{align}
	C_2^f=0 \, &,\quad
	C_4^f=\frac{\epsilon_{ij}^{f}\epsilon_{ji}^{f\, *}}{c_\beta^2\,m_H^2} \, , \quad\quad\quad\quad\quad\ \text{(with decoupling)}\\
	C_2^f=\frac{\epsilon_{ij}^{f\,2}}{4c_\beta^2}\frac{m_A^2-m_H^2}{m_H^2 m_A^2}\, &,\quad
	C_4^f=\frac{\epsilon_{ij}^{f\,}\epsilon_{ji}^{f\, *}}{2c_\beta^2}\frac{m_A^2+m_H^2}{m_H^2 m_A^2} \, , \quad \text{(without decoupling).}
\end{align}
Plugging in these 
expressions in \eqs{eq:boundK}{eq:boundD} and employing the definitions in \eqs{eq:epsilonsum}{eq:epsilonproduct} 
it is straightforward to obtain the meson mixing bounds for the alignment scenarios,   
see \eqs{eq:BoundDecouplingKaon2}{eq:BoundDecouplingD2} %and \eqs{eq:BoundDecouplingKaon2}{eq:BoundDecouplingD2} for the case with and without decoupling, respectively
.  

\section{Model-dependent bounds on the PQ-2HDM}
\label{app:M1M4bounds}

\begin{figure}[t]
    \centering
    \includegraphics[width=\textwidth]{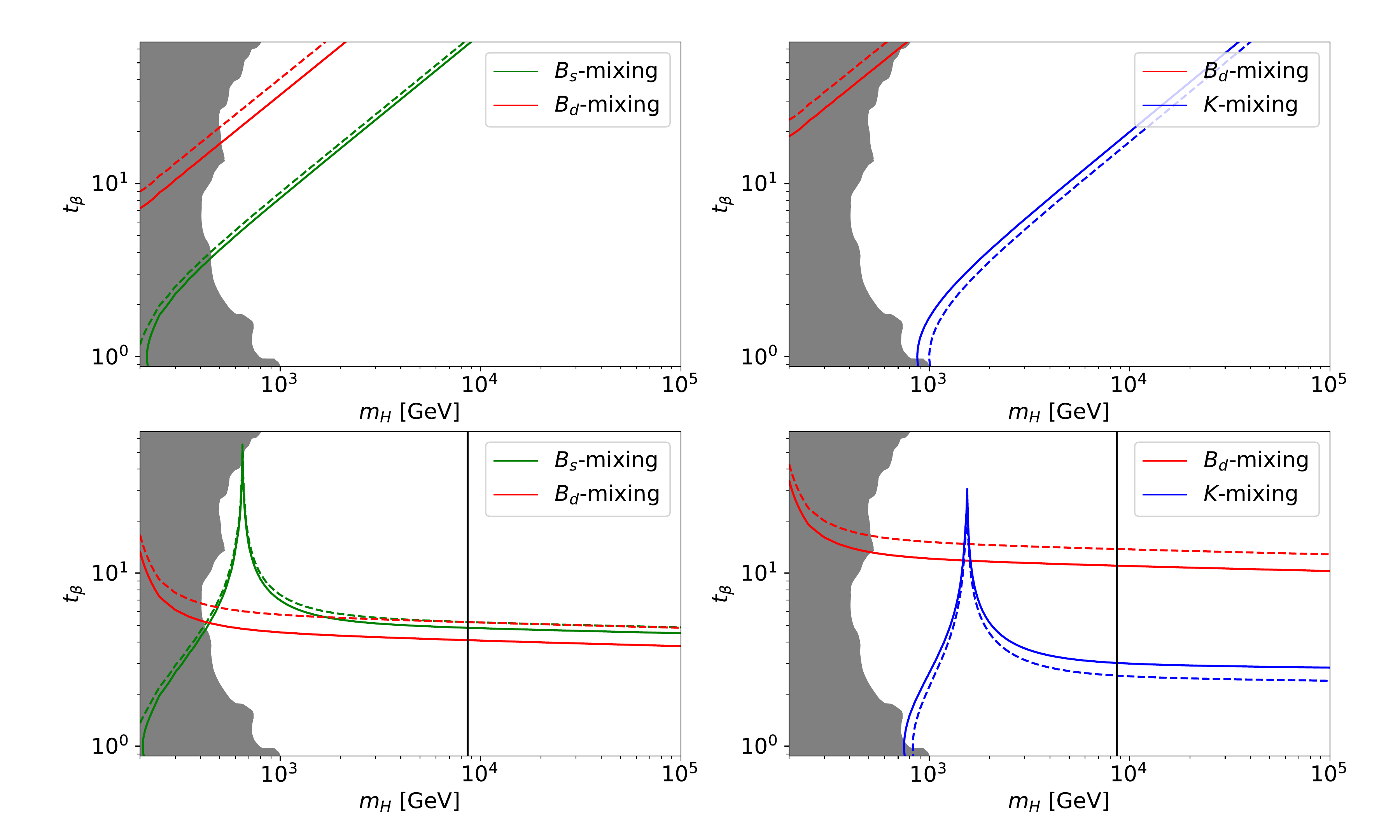}
    \caption{Left (right) panel plots show the meson mixing 
    bounds for the M1 (M4) model, 
    with the solid 
    (dashed)
    line corresponding to a positive (negative) 
    new physics contribution.
    The top panels correspond to 
    $c_{\alpha-\beta}=0$ and the bottom ones to $c_{\alpha-\beta}=0.1$. The black vertical line in the bottom panels indicates the value of $m_H$ for which the 
    condition $c_{\alpha-\beta}=0.1$ saturates the perturbativity of the $\Lambda_1$ parameter, 
    i.e.~$\Lambda_1 \simeq 4\pi$. LHC direct searches 
    limits are displayed as well, 
    taking as a reference those for the case of the type-II 
    2HDM (grey shaded region).}
    \label{fig:m_tb}
\end{figure}
\begin{figure}[t]
    \centering
    \includegraphics[width=\textwidth]{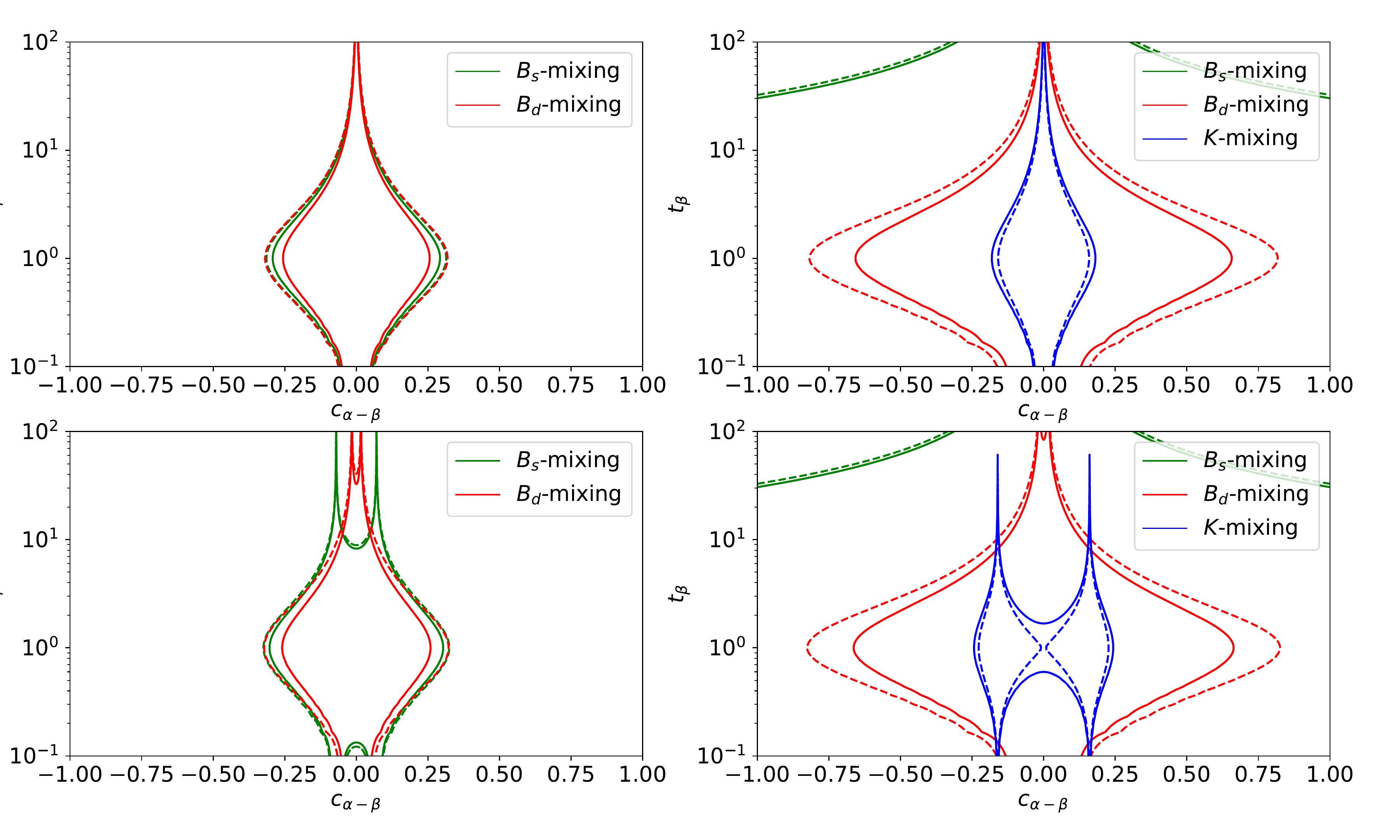}
    \caption{Left (right) panel plots show the meson mixing 
    bounds due to the SM-like Higgs for the M1 (M4) model, 
    with the solid 
    (dashed)
    line corresponding to a positive (negative) 
    new physics contribution.
    The top panels correspond to a decoupled $m_H$
    and the bottom ones to $m_H=1$ TeV.}
    \label{fig:anglesbounds}
\end{figure}

In this Appendix we analyze other aspects of the PQ-2HDM which do not rely on the 
IR/UV connection discussed in this paper, but we assume a certain structure for the PQ charges and the mixing matrices. A possible ansatz is given for instance 
by $V_{d_{L,R}} = V_{\text{CKM}}$ and $V_{u_{L,R}}=\mathbf{1}$.
Considering the M1 and M4 models introduced in \sect{sec:nonuniversalaxion},
we can see from \fig{fig:m_tb} that the bounds become much weaker once we introduce this CKM-like scenario.  
In the M1 model they are around $200\,$GeV. In fact, in this case the 
strongest bound does not come from $K$-mixing, since the PQ charge structure is such that an $s$-$d$ transition is obtained after two CKM rotations of strength $\sim \lambda^3\times \lambda^2$ and hence it is somewhat protected by the 2+1 structure of the PQ charges. In the M4 model instead, 
due to the PQ charge structure, the $K$-mixing leads to the strongest bound, being the $s$-$d$ transition of order $\lambda$.

Note that in the meson mixing bounds (cf.~\eqs{eq:BoundDecouplingKaon2}{eq:BoundDecouplingD2}) 
there could be a cancellation 
for $c_{\alpha-\beta}\neq 0$ (or $m_H \neq m_A$) which correspond to the bottom plots in \fig{fig:m_tb}.
Hence, meson mixing does not rule out a relevant portion of the parameter space. For the M1 the model this is of the order of $200\,$GeV,  
while M4 is still in the reach of future LHC searches. Also, in the eventual discovery of a heavy Higgs, 
if we observe as well flavour mixing, the observed channel could help in reconstructing the flavour structure of the 2HDM.

For completeness, and for comparison with Ref.~\cite{Badziak:2021apn}, 
we also show in \fig{fig:anglesbounds} 
the meson mixing bounds in the $(c_{\alpha-\beta},t_\beta)$ 
plane, for a fixed value of $m_H$. 
In particular, the top panel plots correspond to a mass 
parameter that is large enough to be considered in the 
decoupling regime, while the bottom ones are for 
$m_H=1\,$TeV. 
Here, we display the bound imposed by the SM-like Higgs component contributing to meson mixing. 
Note that it is not possible to decouple
the heavier Higgs without sending to zero the mixing angle of the CP-even states. If we fix the mass of the second Higgs at $1\,$TeV, as in the bottom row of \fig{fig:anglesbounds}, 
then a possible cancellation in the 
meson mixing formula (cf.~\eqs{eq:BoundDecouplingKaon2}{eq:BoundDecouplingD2}) 
is at play. 
This is an interesting parameter space region, since it can be potentially probed by Higgs flavour violating decays, 
such as $h \to b s$.   
As a final comment,
we note that meson mixing constraints in the 
$(c_{\alpha-\beta},t_\beta)$ plane 
are comparable to those obtained from lepton 
flavour violating 
observables (see e.g.~\cite{Badziak:2021apn}).

\section{Lepton flavour violating observables} \label{app:LFVobs}

In this Appendix we provide a compendium of formulae for LFV processes in the parametrization that allows us 
to make a direct connection between the IR and UV sources of flavour violation,
including the 
leptonic decays $l_i\to 3l_j$, $l_i\to l_j l_k l_k$ 
and $l_i\to \gamma l_j$ as well as the LFV Higgs decays 
$h\to l_i l_j$.
Employing the IR/UV relation in \eq{eq:epseC}, 
the contribution of the heavy radial modes 
of the PQ-2HDM to the leptonic decays, $l_i \to 3l_j$, 
can be written as \cite{Crivellin:2013wna}
\begin{align}
	\text{BR}(l_i\to 3l_j)&=\frac{m_\tau^5}{96\pi^3 \Gamma_\tau\, m_H^4}\(\frac{m_j}{s_{2\beta} v}\)^4(C^A_e)_{jj}^2 \nonumber \\
&\times  
 \[\left|(C^R_e)_{ji}+\frac{m_\tau}{m_j}(C^L_e)_{ji}\right|^2 +\left|(C^L_e)_{ji}+\frac{m_\tau}{m_e}(C^R_e)_{ji}\right|^2 \] \, ,
\end{align}
where we have neglected the mass of the final state lepton. This equation contains a combination of diagonal and off-diagonal components, thus making the IR/UV connection 
a bit more involved.\footnote{In this case 
one could fix the diagonal axion coupling to electrons 
to the value suggested by the cooling hints \cite{DiLuzio:2021ysg}, and then impose the LFV bounds 
in analogy to what done in \sect{sec:IRUV_meson} for the 
quark sector.} 
Again, here we consider $f_a\gg m_H$, so that the 
direct axion 
contribution to this observable is negligible. 

Within 2+1 PQ models 
(featuring axion flavour violation either in the LH or RH sector)
the expression above 
gets further simplified into
 \beq
     \label{eq:lito3lj}
 	\text{BR}(l_i\to 3l_j)= \frac{m_{f_i}^5}{96\pi^3 \Gamma_i\, m_H^4}\(\frac{\sqrt{m_{f_j}m_{f_i}}}{s_{2\beta} v}\)^4(C^A_e)_{jj}^2 \left|(C^{R,L}_e)_{ji}\right|^2 \, .
 \eeq
For $l_i \to l_j \bar l_k l_k$ 
decays 
we obtain instead
(always in the case of 2+1 PQ models) 
 \begin{align}
 	\text{BR}(l_i\to l_j\bar l_k l_k)&=\frac{m_{f_i}^5}{96\pi^3 \Gamma_i\, m_H^4}\(\frac{m_{f_i}}{s_{2\beta} v}\)^4\[(C^A_e)_{kk}^2 \left|(C^{R,L}_e)_{ji}\right|^2\frac{m_k^2}{m_i^2}\(1+\frac{m_j^2}{m_i^2}\) \right. \nonumber \\&+ \left. \left|(C^{R,L}_e)_{jk}\right|^2 \left|(C^{R,L}_e)_{ki}\right|^2\(\frac{m_k^4}{m_i^4}+\frac{m_j^2}{m_i^2}\)\] \, .
 \end{align}
Following \cite{Crivellin:2013wna} we can also write the 
one-loop contribution of the PQ-2HDM to the radiative 
LFV
decays 
as 
\beq
     \text{BR}(l_i\to l_j \gamma)=\frac{m_{l_i}^5}{4\pi\Gamma_{l_i}}\(|C_R^{l_j,l_i}|^2
     +|C_L^{l_j,l_i}|^2\) \, ,
\eeq
in terms of the coefficients $C_{L(R)}^{l_j,l_i} = C_{L(R),+}+C_{L(R),0}$, denoting the sum of the charged and neutral contributions. 
For simplicity we will give them 
here for the case 
of 2+1 PQ models. The charged contribution reads
\begin{align}
    C_{R,+}^{l_j,l_i}=\frac{e}{384\pi^2m_{H^\pm}^2}\left| C^{H^+_e}\right|^2_{ji}\, , \qquad C_{L,+}^{l_j,l_i}=\frac{e}{384\pi^2 m_{H^\pm}^2}\frac{m_{l_j}}{m_{l_i}}\left| C^{H^+_e}\right|^2_{ji} \, , 
\end{align}
which we did not expand in terms of the axion couplings, 
since it includes both diagonal and off-diagonal couplings. The neutral contribution reads instead 
(working in the decoupling limit, $m_H^2 \simeq m_A^2$)
\begin{align}
	 C_{L(R),0}^{l_j,l_i}&=\frac{-e}{96 \pi^2 \, m_H^2}\(\frac{\sqrt{m_{e_i} m_{e_j}}}{v c_\beta s_\beta}\)^2 
  \Bigg[  2(C^A_e)_{ii}(C^{L(R)}_e)_{ji} +\(1+\frac{m_{e_j}^2}{m_{e_i}^2}\)(C^A_e)_{jj}(C^{L(R)}_e)_{ji} \nonumber \\ &-\(1+\frac{m_{e_k}^2}{m_{e_i}^2}\)(C^{L(R)}_e)_{jk}(C^{L\,(R)}_e)^*_{ik}\Bigg] \, , \\
	 C_{R(L),0}^{l_j,l_i}&=\frac{e}{96 \pi^2 \, m_H^2 }\(\frac{\sqrt{m_{e_i} m_{e_j}}}{v c_\beta s_\beta}\)^2\frac{m_{e_j}}{m_{e_i}} 
  \Bigg[  2(C^A_e)_{jj}(C^{L(R)}_e)_{ji}+\(1+\frac{m_{e_i}^2}{m_{e_j}^2}\)(C^A_e)_{ii}(C^{L(R)}_e)_{ji}  \nonumber \\ 
  &-\(1+\frac{m_{e_k}^2}{m_{e_j}^2}\)(C^{L(R)}_e)_{jk}(C^{L(R)}_e)^*_{ik}\Bigg] \, ,
\end{align}
where $k\neq i,\, j$ is not summed over. 
Note that due to the structure of the loop contribution, 
it is not possible to obtain a straightforward 
connection with LFV axion observables.   
However, the bounds from radiative LFV decays 
such as $\mu\to e\gamma$ 
are 
very relevant  
for constraining the parameter space
and they should be 
eventually taken into acccount in a global analysis.

We finally consider LFV Higgs decays, 
whose connection with LFV axion processes was discussed 
in \sect{sec:LFV_connection}. 
In this Appendix we provide full formulae, 
including the effects of a modified Higgs width 
and show that also flavour conserving Higgs decays can be related to axion physics. The complete formula for the 
LFV Higgs decays can be written as 
(in terms of LFV axion couplings, exploiting \eq{eq:epseC})
\begin{align}
\label{eq:generalhLFV}
    \text{BR}(h\to l_i l_j)&=\frac{m_h}{16\pi\Gamma_h}\( \frac{c_{\alpha-\beta}}{v s_\beta c_\beta}\)^2
    \Big[\( m_{l_i}^2+m_{l_j}^2\)\(|(C^L_e)_{ij}|^2 +|(C^R_e)_{ij}|^2\) \nonumber \\ &-4m_{l_i} m_{l_j}\Re{\((C^L_e)_{ij}(C^R_{e})^*_{ij}\)}\Big] \, , 
\end{align}
where the Higgs width is modified as \cite{Badziak:2021apn}
\beq
    \label{eq:ModHiggsWidth}
 	\Gamma_h\simeq \frac{k_h^2}{1-B^{\text{BSM}}_h}\Gamma_{\text{SM}} \, , 
\eeq
with $B^{\text{BSM}}_h$ denoting the branching ratio 
of the beyond-the-SM contributions to the Higgs decay width, $\Gamma^{\text{SM}}_h \simeq 4.1\,$MeV and $k_h^2$ 
parametrizing the effect due to Higgs couplings modifications. The latter is defined 
as \cite{Badziak:2021apn}
\begin{align}
 	k_h^2 &\simeq 0.58 k_b^2+0.22 k_W^2+0.08 k_g^2+0.06k_\tau^2+0.03 k_Z^2+0.03 k_c^2+2.3\times10^{-3}k_\gamma^2 \nonumber \\
 	&+1.5\times 10^{-3}k_{Z\gamma}^2+4\times10^{-4}k_s^2+2.2\times10^{-4}k_\mu^2 \, , 
\end{align}
in terms of the $k$-parameters 
\begin{equation}
    k_{f_i}\equiv \frac{\sqrt{2}m_f}{v}C^{h_f}_{ii}=-s_{\alpha-\beta}+\frac{c_{\alpha-\beta}}{c_\beta s_\beta}(C^A_{f})_{ii} \, , 
\end{equation}
where in the last step we have exploited again \eq{eq:epseC}. 
In particular, the modified Higgs interactions with gauge couplings read
\begin{align}
 	&k_W=k_Z=s_{\alpha-\beta}\, , \\
    &k^2_{Z\gamma}\simeq0.00348k_t^2+1.121k_W^2-0.1249k_tk_W\, , \\
 	&k_g^2 \simeq 1.04k_t^2+0.002k_b^2-0.04k_b k_t\, , \\
 	&k_\gamma^2 \simeq 1.59 k_W^2+0.07k_t^2-0.67k_W k_t \, .
\end{align}
Hence, the above expressions show that deviations in Higgs couplings can also be related to diagonal axion couplings. 

\begin{small}

\bibliographystyle{utphys}
\bibliography{bibliography.bib}

\end{small}

\clearpage

\end{document}